\documentclass[fleqn,10pt]{wlscirep}

\RequirePackage{xcolor}
\usepackage{caption}
\usepackage{graphicx}
\usepackage{color, colortbl}
\definecolor{darkGray}{gray}{0.9}
\definecolor{lightGray}{gray}{0.6}

\title{Parsing spatiotemporal dynamical stability in ECoG during seizure onset, propagation, and termination}

\author[1,6,*]{Arian Ashourvan}
\author[2,*]{S\'ergio Pequito}
\author[1,3]{Ankit N. Khambhati}
\author[1,3]{Steven N. Baldassano}
\author[3,4]{Kathryn A. Davis}
\author[3,5]{Timothy Lucas}
\author[6,7]{Jean M. Vettel}
\author[1,3,4]{Brian Litt}
\author[2]{George J. Pappas}
\author[1,3,$\dag$]{Danielle S. Bassett}

\affil[*]{Both authors contributed equally to this work.}
\affil[1]{Department of Bioengineering, School of Engineering and Applied Science, University of Pennsylvania, Philadelphia, PA 19104 USA}
\affil[2]{Department of Electrical and Systems Engineering, School of Engineering and Applied Science, University of Pennsylvania, Philadelphia, PA 19104 USA}
\affil[3]{Penn Center for Neuroengineering and Therapeutics, University of Pennsylvania, Philadelphia, PA 19104 USA}
\affil[4]{Department of Neurology, Hospital of the University of Pennsylvania, Philadelphia, PA 19104 USA}
\affil[5]{Department of Neurosurgery, Hospital of the University of Pennsylvania, Philadelphia, PA 19104 USA}
\affil[6]{U.S. Army Research Laboratory, Aberdeen Proving Ground, MD 21005 USA}
\affil[7]{Department of Psychological \& Brain Sciences, University of California, Santa Barbara, CA, 93106 USA}

\affil[$\dag$]{Correspondence to: Danielle S. Bassett, 210 S. 33rd Street, Philadelphia, PA 19104-6321,  Email:~dsb@seas.upenn.edu}

\begin{abstract}
Understanding brain dynamics in epilepsy is critical for establishing rigorous control objectives that enable new therapeutic methods to mitigate seizure occurrence. In multichannel electrocorticography (ECoG) recordings acquired in 21 subjects during a total of 94 seizures, we apply dynamical systems stability analysis to assess the balance \emph{versus} imbalance of seizure dynamics across different timescales and brain regions. Specifically, we consider a sliding time window multivariate autoregressive linear approximation of the data captured by the ECoG channels, where eigendecomposition of the estimated matrix of coefficients describes the contribution of different regions to the spatiotemporal process (eigenvectors) associated with a particular timescale (eigenvalues). Interestingly, we observe a pattern of eigenvalue evolution and slowly changing (or approximately time-invariant) eigenvectors across both seizures and subjects. The seizure-onset is marked by an increase in high frequency spatial information to which a few regions contribute for a long period. By contrast, the seizure termination is characterized by a sudden, small time period change in dynamics to which many regions contribute. As the seizure terminates, the relatively stable ictal dynamics rapidly transition into the post-ictal regime, marked by relatively fast-damping oscillations. Our methodology offers a subject-specific characterization of the spatiotemporal behavior of the seizure, providing new insights into the dynamic patterns and functional interactions between brain regions that occur over different timescales. More generally, our approach informs the development of engineering objectives that can be used to deploy new control strategies to prevent seizure evolution or to hasten seizure termination. 
\end{abstract}
\begin{document}

\flushbottom
\maketitle

\thispagestyle{empty}


\section*{Introduction}

Understanding the dynamic nature of the neurophysiological processes underlying seizure initiation is critical for developing insight into epileptogenesis, seizure evolution, and seizure termination \citep{pitkanen2016advances}. Fundamental knowledge about these processes could augment burgeoning neurotechnologies, such as implantable devices \citep{Morrell2011}, by informing patient-centric algorithms that modulate brain dynamics to abort seizures \citep{Stacey2008, Stanslaski2012, Afshar2013}. Ideally, such an algorithm would account for both \emph{when} and \emph{where} to deliver a perturbative payload within the brain's physiological network. Indeed, spatiotemporal precision in stimulation protocols could improve quality of life for the nearly 1\% of the world's population affected by epilepsy \citep{mormann2007seizure}. Advances in stimulation protocols may be facilitated by theoretically-driven principles from engineering and control theory \citep{schiff2012neural}, which can inform interventions to mitigate the physiological effects of ictal activity \citep{schindler2007assessing}. Progress will critically depend on building a deeper understanding of the types of dynamical states that seizures represent in the context of broader neural systems. 

Over the past decade or more, an important body of work has sought to connect seizure physiology to dynamical systems theory \citep{schindler2007assessing, Rummel2013}. Several exploratory tools have been developed to identify features in EEG and ECoG signals that are associated with ictal processes and can serve as biomarkers for seizure detection and prediction \citep{Baldassano2017Crowdsourcing, brinkmann2016crowdsourcing}. These features are identified either by concepts in non-parametric statistics and signal processing, or by multivariate analysis of parametric models \citep{schomer2012niedermeyer}. The former approach can provide classifications of seizure activity and seizure-onset types based on analysis of EEG or ECoG recordings \citep{perucca2014intracranial}. The latter approach relies on parametric models and has the added benefit of potentially capturing the substrate of the dynamic process itself -- reflected in spatiotemporal properties of the EEG or ECoG activity \citep{schindler2007assessing, schiff2005neuronal, Nair2004, Kramer2010, Warren2010, Wilke2011, Burns2014, Khambhati2015}. The development of more accurate predictors and more physiologically interpretable models is an important and ongoing area of research.

These efforts are particularly challenging because epileptiform activity manifests across a range of temporal and spatial scales and displays marked inter-channel dependencies \citep{jirsa2014nature}. Understanding the biological phenotype therefore requires the ability to assess interactions between different channels across different timescales, with the potential of providing testable causal mechanisms of seizure generation, progression and termination. To address this challenge, we exercise a paradigm that relies on the concept of dynamical systems stability to study the behavior of neural processes in the vicinity of a putative bifurcation between dynamically stable and unstable behaviors \citep{magnasco2009self}. The approach is built on the notion that seizures can occur as a result of a dysfunction in regulatory mechanisms that might otherwise maintain stable brain dynamics \citep{koepp2016reflex}.  Our approach provides a compact representation and an intuitive visual mapping that captures the interactions between channels across different time-scales -- a capability that can potentially provide a dynamics-based interpretation of epileptic neural signals in clinical neuroscience. Further, our approach offers a dynamical definition of the onset and termination phases of a seizure, a critical step toward understanding the natural processes that govern seizure dynamics \citep{schiff2005neuronal}. 

 We apply this dynamical stability analysis approach to ECoG data acquired from twenty one subjects undergoing surgical treatment for medically refractory epilepsy that is likely of neocortical origin. Our results suggest that assessments based on signal frequency alone may miss salient changes in the spatio-temporal evolution of the dynamical interaction processes characteristic of epileptiform activity. Moreover, we demonstrate that spatio-temporal disruptions in dynamical stability mark periods of both seizure onset and propagation. Our data suggest that inter-ictal dynamics are an emergent macroscopic phenomenon characterized by coherent activity among different cortical regions, whereas the ictal regime is characterized by the emergence of focal and persisting dynamics. Finally, we note that the proposed approach provides us with a criterion that is useful for future efforts in stimulation control, supporting the development of new stimulation strategies to be used in therapeutic implantable devices \citep{osorio2005automated,kim2009dynamics,echauz2009intelligent,morrell2006brain}.

\section*{Materials and Methods}

\subsection*{Description of experimental data}

In this study, we considered twenty one subjects undergoing surgical treatment for medically refractory epilepsy, likely of neocortical origin. These subjects underwent the implantation of subdural electrodes to assess seizure location and evolution after noninvasive monitoring proved inconclusive. All data had previously been de-identified and was publicly available via the International Epilepsy Electrophysiology Portal (IEEG Portal) ~\citep{IEEG2013}. Patients exhibited a range of seizure etiology, location, type, and degree of severity, providing a rich dataset for the assessment of seizure dynamics. In Table~\ref{tab:dataset}, we report important details about the subjects that were monitored at the Hospital of the University of Pennsylvania (Philadelphia, PA, USA) and at the Mayo Clinic (Rochester, MN, USA); subjects from these two cohorts were labeled as HUP and Study, respectively. Specifically, in Table~\ref{tab:dataset}, we report the gender (Sex) as well as the age at first reported onset and at phase~II monitoring (Age). Additionally, we report the localization of seizure-onset and the seizure etiology, which was clinically determined through medical history, imaging, and long-term invasive monitoring. The different seizures observed (Seizure Types) included simple-partial (SP), complex-partial (CP), and complex-partial with secondary generalization (CP+GTC). In this table, we also indicate the total number of seizures recorded in the epilepsy monitoring unit, as well as the clinical imaging analysis (Imaging) that concludes if the seizure etiology is lesional (L) or non-lesional (NL). Finally, surgical outcome (Outcome) was based on either Engel score or ILAE score: seizure freedom to no improvement (I-V), no resection (NR), and no follow-up (NF). 

\begin{table}[ht]
\centering
\resizebox{\textwidth}{!}{
  \begin{tabular}{cccccccccc}
  \hline \begin{tabular}{@{}c@{}} \textbf{Subject}\\  (\#)\end{tabular} & \begin{tabular}{@{}c@{}}\textbf{Patient} \\ (IEEG Portal) \end{tabular}
 & \textbf{Sex}  & \begin{tabular}{@{}c@{}} \textbf{Age} (Years) \\ (Onset/Surgery) \end{tabular}  &\textbf{seizure-onset} & \textbf{Etiology} & \textbf{Seizure Type} & \begin{tabular}{@{}c@{}} \textbf{Seizures}\\  (\#)\end{tabular} & \textbf{Imaging} & \textbf{Outcome} \vspace{0.1cm}\\ 
\rowcolor{darkGray}
1&HUP64\_phaseII  &M& 03/20 & Left frontal &Dysplasia& CP+GTC &01& L& ENGEL-I\\
2&HUP65\_phaseII &M& 02/36 &Right temporal & Unknown &CP+GTC &03 &N/A &ENGEL-I\\
\rowcolor{darkGray}
3&HUP68\_phaseII &F& 15/26 &Right temporal &Meningitis &CP, CP+GTC & 05 & NL &ENGEL-I \\
4&HUP70\_phaseII &M& 10/32 & Left perirolandic &Cryptogenic &SP &08& L &NR\\
\rowcolor{darkGray}
5&HUP72\_phaseII &F& 11/27 & left frontal & Unknown &CP+GTC &01 & NL \\
6&HUP73\_phaseII &M& 11/39 &Anterior right frontal &Meningitis & CP & 05& NL&  ENGEL-I\\
\rowcolor{darkGray}
7&HUP78\_phaseII &M& 00/54 &Anterior left temporal &Traumatic brain injury &CP& 05& L& ENGEL-III\\
8&HUP79\_phaseII &F& 11/39 & Occipital & Meningitis& CP& 02& L& NR\\
\rowcolor{darkGray}
9&HUP86\_phaseII &F& 18/25 & Left temporal &Cryptogenic &CP+GTC &02& NL& ENGEL-II\\
10&HUP87\_phaseII &M& 21/24 & Frontal & Meningitis &CP & 02& L& ENGEL-I\\
\rowcolor{darkGray}
11&Study 006 &M& 22/25 & Left frontal &Unknown &CP& 02 &NL &NR\\
12&Study 010 &F& 00/13 & Left frontal & Neonatal injury &CP+GTC& 02 &L &NF\\
\rowcolor{darkGray}
13&Study 011 &F& 10/34 & Right Mesial frontal &Unknown & CP & 02 & NL &NF\\
14&Study 016 &F& 05/36 & Right temporal orbitofrontal &Unknown& CP, CP+GTC& 04& NL& ILAE-IV\\
\rowcolor{darkGray}
15&Study 019 &F& 31/33 & Left temporal &Unknown&  CP, CP+GTC& 19& NL& ILAE-V\\
16&Study 020 &M& 05/10 & Right frontal &Unknown& CP& 05& NL& ILAE-IV\\
\rowcolor{darkGray}
17&Study 023 &M& 01/16 & Left occipital & Traumatic brain injury& CP& 04& L& ILAE-I\\
18&Study 026 &M& 09/09 & Left frontal &Unknown& CP& 10& NL& ILAE-I\\
\rowcolor{darkGray}
19&Study 031 &M& 05/05 & Right frontal & Focal cortical dysplasia taylor type Iia & CP+GTC& 05& NL& NF\\
20&Study 033 &M& 00/03  & Left frontal &Tuberous Sclerosis & CP & 02& L& ILAE-V\\
\rowcolor{darkGray}
21&Study 037 &F& 45/62  & Indeterminate &Unknown& CP,  CP+GTC & 05& NL& NR\\
\end{tabular}
 }

 \caption{\textbf{Patient Information}. For each patient, we report gender (Sex), and age at first reported onset and at phase~II monitoring (Age). Additionally, we report the localization of seizure-onset and the seizure etiology, which was clinically-determined through medical history, imaging, and long-term invasive monitoring. The different seizures observed (Seizure Types) include simple-partial (SP), complex-partial (CP), and complex-partial with secondary generalization (CP+GTC). We also indicate the total number of seizures recorded in the epilepsy monitoring unit, as well as the clinical imaging analysis (Imaging) that concludes if the seizure etiology is lesional (L) or non-lesional (NL). Finally, surgical outcome (Outcome) was based on either Engel score or ILAE score: seizure freedom to no improvement (I-V), no resection (NR), and no follow-up (NF). } 
 \label{tab:dataset}
\end{table}

\subsection*{Intracranial electroencephalogram recordings}

ECoG signals were recorded and digitized at sampling rates of 512 Hz (Hospital of the University of Pennsylvania, Philadelphia, PA) and 500 Hz (Mayo Clinic, Rochester, MN). Surface electrode (Ad Tech Medical Instruments, Racine, WI) configurations, determined by a multidisciplinary team of neurologists and neurosurgeons, consisted of linear and two-dimensional arrays (2.3 mm diameter with 10 mm inter-contact spacing) and sampled the neocortex for epileptic foci (depth electrodes were first verified as being outside the seizure-onset zone and subsequently discarded from this analysis). Signals were recorded using a referential montage with the reference electrode, chosen by the clinical team, distant to the site of seizure-onset. Recording spanned the duration of a patient's stay in the epilepsy monitoring unit. 

A total of 94 seizure-onsets were determined and the seizure-onset time and localization were defined by the point of ``earliest electrographic change'' (EEC), which was annotated and marked by a team of practicing epileptologists \citep{litt2001epileptic}. The EEC associated with a given seizure is found by moving backward in time from the unequivocal electrographic onset (UEO) until the earliest change from previous background activity can be observed visually on the ECoG traces. The UEO is defined as the first unequivocal intracranial EEG sign of change from the background, which leads to a clear seizure discharge without return to background activity \citep{spanedda1997relations}. Using the EEC, we extracted multi-channel ECoG time series from seizure onset to seizure termination. In addition, we also extracted ECoG time series 20 seconds directly preceding (pre-ictal) and following (post-ictal) each seizure. Lastly, we extracted a total of 478 (22.7 $\pm 7.3$ segment per patient) randomly-selected inter-ictal segments (lasting 100 seconds each) from traces acquired more than $3$ hours before or $3$ hours after a seizure. All time series were high-pass filtered to maintain information in the frequency range $>0.1$ Hz.

\subsection*{Dynamical stability approach}

ECoG signals can be locally approximated at each time point by a linear system \citep{khalil2014nonlinear}. Let $x(k)\in \mathbb{R}^n$ be the data from $n$ channels, where the $i$th entry $x_i(k)$ corresponds to the data collected by channel $i$ at time $k$. Then, we obtain
\begin{equation}
x(k)=Ax(k-1)+\varepsilon(k),
\label{linearDyn}
\end{equation}
where $A$ is the $n\times n$ real matrix that results from fitting the data collected within the interval of time $[k-\tau,k+\tau]$ using a least squares approach, and where $\varepsilon(k)$ is the approximation error. As a consequence of the linear approximation, the results will depend on the choice of the number $n$ of channels and the size of the time window parametrized by $\tau$. We will discuss the impact of these parameters at greater length in the Results section and in the Supplementary Materials document.

A significant advantage of using a linear approximation of sensor dynamics is that the dynamical properties of the underlying process can be locally assessed through the eigendecomposition of $A$: that is, the $n$ eigenvalue-eigenvector pairs. In other words, eigenvalue-eigenvector pairs capture linearly independent spatiotemporal dynamical processes. Specifically, for each putative spatiotemporal process, the eigenvector weights the involvement of each sensor and the complex eigenvalue defines the frequency of the oscillatory dynamics. More importantly, the stability of the dynamics can be captured by the absolute value of the eigenvalues, which forecast the exponential growth or decay along the associated eigenvector. 

Specifically, let $A=V\lambda V^{\intercal}$ be the eigendecomposition for a given time point $k$, where $V=[v_1,\ldots, v_n]$ and $\lambda=\text{diag}(\lambda_1,\ldots,\lambda_n)$ are the matrices of eigenvectors and eigenvalues (when the eigenvalues are all distinct complex numbers) respectively, where $(\lambda_i,v_i)$ are the eigenvalues paired with the associated eigenvector, and where $V^{\intercal}$ is the transpose of $V$. Notice that some eigenvalues are complex numbers, which implies that no partial-order can be imposed on the eigendecomposition. Further, note that after $T$ time steps, one obtains from Eq.~\eqref{linearDyn} that $x(k)=A^{T}x(0)$, which implies that $x(k)=V\lambda^T V^{\intercal}x(0)$. Subsequently, by considering a linear combination of the original data $z(k)=V^*x(k)$, where $z_i(k)=v_i^\intercal x(k)$ is a weighted combination described by the $i$th eigenvector associated with the $i$th eigenvalue. 

From these definitions, it follows that $|z_i(k)|=|\lambda_i|^T|z_i(0)|$ and three scenarios are possible: (\emph{i})  $|\lambda_i|< 1$ which leads to $|z_i(k)|\rightarrow 0$ as $t\rightarrow \infty$; (\emph{ii}) $|\lambda_i|> 1$ which implies that $|z_i(k)|\rightarrow \infty$ as $t\rightarrow \infty$; and $|\lambda_i|= 1$ where $|z_i(k)|= |z_i(0)|$ for all times. In (\emph{i}), the process tends to vanish, and we therefore refer to these dynamics as \emph{asymptotically stable}. In (\emph{ii}), the process tends to explode, and we therefore refer to these dynamics as \emph{unstable}. Finally in (\emph{iii}), the process oscillates between stability and instability, and we therefore refer to these dynamics as \emph{stable}. In practice, we consider that the stable regime is determined by $|\lambda_i|\approx 1$. Since these processes are associated with~$z(k)$, it follows that they are associated with a specific eigenvector, and, consequently, we refer to a stable regime associated with a particular eigenvector. Thus, the interplay between these three different stages and different eigenvectors provides a \emph{dynamical stability} characterization to identify seizure onset and to monitor seizure evolution.  
 
To further characterize these dynamics, we note that the angle $\theta_i$ associated with the polar coordinates of the $i$th complex eigenvalue provides a description of the \emph{frequency} as follows:
\[
f_i=\frac{\theta_i}{2\pi}\delta t,
\]
where $\delta t$ corresponds to the sampling frequency; the \emph{timescale} is given by
\[
\rho_i=\frac{\log(|\lambda_i|)}{\delta t}
\]
\noindent which can be interpreted as the growth rate. Therefore, the dynamical process $z(k)$ describes the spatiotemporal behavior of the dynamical system, where the timescale is encoded in the eigenvalue and the spatial scale is encoded in the eigenvector, indicating the relative contribution of a given channel or -- by extension -- the overall activity in a specific cortical region. 

In the main text, we use this approach to consider the time-evolution of \emph{eigenvalue-eigenvector} pairs associated with higher and lower frequencies, and associated with fine and coarse timescales, to better understand seizure onset, propagation, and termination. In the Supplementary Materials, we define and study synthetic examples of damping \emph{versus} growing oscillations and build the reader's intuition for how dynamical stability analysis allows us to understand and quantitatively characterize the underlying process.

\subsection*{Statistical testing}

Because the functional forms of the distributions of our statistics of interest were not known \emph{a priori}, we applied non-parametric permutation testing to assess statistical significance of our findings. 

As we will describe in greater detail in the Results section, we first sought to assess the extent of the time-invariance of the eigenvectors over the ictal period. We began by defining a summary statistic of interest. Specifically, we computed the sum of the element-wise squared differences between temporally contiguous eigenvectors; then, we calculated the average standard deviation in this value across all temporally contiguous pairs of eigenvectors across an ictal event. Intuitively, this summary statistic quantifies the temporal variance of the given eigenvector across time: high values of this statistic indicate large temporal fluctuations, and small values of this statistic indicate small temporal fluctuations. 

Next, we assessed the frequency-dependence of this statistic at the group level by clustering eigenvectors into five non-overlapping frequency bands based on the eigenvectors' average frequencies (across independent model fits over the time windowed data) over each sample (see Supplementary Materials Fig.~\ref{fig:New_SI_Fig_C} and ~\ref{fig:New_SI_Fig_H} and associated text for further details). We calculated subject-level summary statistics, averaged over all eigenvectors in each band, for (i) the ictal samples, averaged over all samples for each patient, and (ii) the inter-ictal samples, averaged over all samples for each patient. Then, we compared these summary statistics between ictal and inter-ictal samples at the group-level via bootstrap analysis. To create non-parametric null distributions of the group-level average of the difference between the ictal and inter-ictal summary statistics computed at the subject-level, we permuted the labels ``ictal'' and ``inter-ictal'' uniformly at random 50,000 times. Finally, we tested the hypothesis that the eigenvectors in the ictal samples at each frequency band on average displayed significantly less temporal fluctuations than inter-ictal samples across subjects. 

Following the examination of the time-dependence \emph{versus} time-invariance of eigenvectors during the ictal period, and the frequency dependence of our findings as a function of ictal \emph{versus} inter-ictal samples, we finally turned to an assessment of the regional localization of the effects. To measure regional localization, we studied the absolute magnitude of the maximally contributing elements of the eigenvectors. Then, we used a bootstrap analysis to test the hypothesis that the emergence of the slow-changing focal ictal dynamics was associated with an increase in the absolute magnitude of the maximally contributing elements of eigenvectors, particularly within a short time window (1, 3, 6, and 9 s) following the seizure onset. To determine the specificity of the effects to the ictal period, we compared the subject-level summary statics in the ictal segments to those estimated from the pre-ictal segments (see Supplementary Materials Fig.~\ref{fig:New_SI_Fig_I} and associated text for further details).

\section*{Results}

In ECoG traces obtained from twenty-one patients with medically refractory epilepsy, we used dynamical stability analysis to estimate Eq.~\eqref{linearDyn} over a 1 s sliding-window for 100 ms shifts. Specifically, we estimated the matrices $\mathbf{A}$ using previously developed computational algorithms \citep{neumaier2001estimation}, and then we calculated the \emph{eigenmodes} or eigenvalue-eigenvector pairs of the estimated $\mathbf{A}$ matrices for each time point $k$. The eigenvectors are considered on the basis of their spatiotemporal frequency: that is, the frequency determined by the corresponding eigenvalue. Because the eigenvector may contain several non-zero entries, the frequency of the dynamic process along the direction determined by the eigenvector does not coincide with any specific ECoG channel. Instead, it corresponds to the spatiotemporal frequency along the direction captured by the eigenvector. For ease of visualization, we normalize the eigenvectors by dividing each entry by the maximum value obtained across the entries, and then we take the absolute value. Thus, all entries correspond to values between~$0$ and~$1$, and capture the contribution of different ECoG channels to the evolution of the underlying neurophysiological process at a certain spatiotemporal frequency.

\subsection*{Dynamical stability analysis pinpoints seizure onset by distinct spatiotemporal patterns}

Based on our preliminary observation of the sustained and focal nature of the ictal onset activity, we hypothesized that eigenmodes would evolve with slow spatio-temporal variance, characterized by eigenvectors that remained relatively consistent in their composition and magnitude over a period of time. Moreover, we anticipated distinct phases of a seizure could be distinguished based on their \emph{frequency} as given by the angle associated with the eigenvalues, and based on their \emph{stability} as measured by the radius associated with the eigenvalues. To examine these hypotheses, we first consider the phenotypes in a single subject, and then we examine the consistency of our results across the population.

\begin{figure}[p]
\centering
\includegraphics[width=0.85 \linewidth]{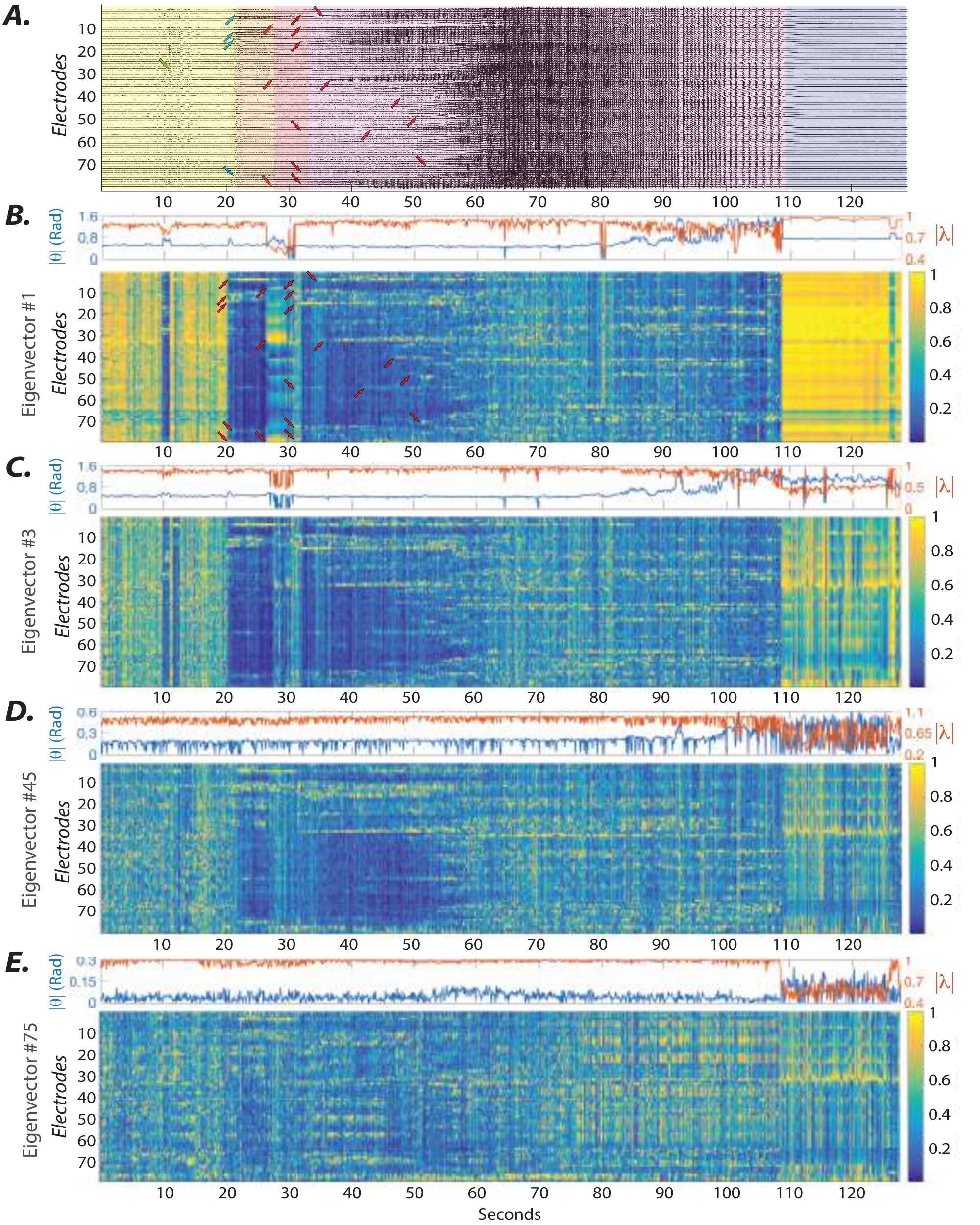}
\caption{\textbf{Evolution of eigenmodes over the ictal period.} \emph{\textbf{(A)}} ECoG signals from 79 electrodes in subject 3 over a single ictal period, from 20 s before seizure onset to 20 s after seizure termination. Pre-ictal activity (yellow), earliest electrographic change (orange), unequivocal seizure-onset (pink), ictal period (purple), and post-ictal period (gray) are delineated separately. \emph{\textbf{(B-E)}} The frequency (cyan trace) and stability (red trace) of eigenvalues associated with four representative eigenvectors -- two associated with high frequencies and two associated with low-frequencies -- whose evolution is displayed in heatmaps. Here we observe that the onset of ictal activity is marked by the emergence of a period of approximate spatiotemporal-invariance in the loading patterns on eigenvectors associated with high frequency dynamics (see approximate spatiotemporal invariance from 20-27 s and from 27-32 s in \emph{\textbf{(B)}}). We also observe an increase in the angle of the eigenvalues time-locked to the onset of these periods. Pointer arrows in panels \emph{\textbf{(A)}} and \emph{\textbf{(B)}} are offered as guides to the eye when comparing epileptiform activity in the ECoG traces and in the eigenvector associated with the highest frequency dynamics. } 
\label{fig:Figure1}
\end{figure}

We begin by considering a randomly selected seizure from a single subject in our sample; this seizure is a complex-partial seizure with secondary generalization, as can be seen in the ECoG traces presented in Fig.~\ref{fig:Figure1}A. As the seizure begins, we observe the emergence of a sustianed regime whereby the eigenvectors associated with high frequency oscillatory dynamics display approximate time-invariance over several seconds (Fig.~\ref{fig:Figure1}B-C: period between 20-27 s). Next, we observe that the eigenvector associated with the highest frequency (eigenvector 1) has high magnitude elements, in a small set of cortical regions: electrodes 4, 11,14, 74, and 79 annotated by red arrows in Fig.~\ref{fig:Figure1}B at the period between 20-27 s. (For a thorough comparison between these marking and clinical markings, please see the supplementary materials section ``Comparison to clinical annotations''.) This point characterized by regional specificity of the eigenvector loadings, as well as approximate time-invariance of the eigenvectors marks seizure initiation. Following this point, we observe that the eigenvector associated with the highest frequency tracks the emergence of asymptotically stable behavior (i.e., fast-dampening oscillations), and again displays a period of approximate time-invariance in eigenvector loading patterns (Fig.~\ref{fig:Figure1}B: period between 27-33 s). We note that the loadings onto the eigenvector in this period are distinct from that in the previous seizure initiation period.

\begin{figure}[p]
\centering
\includegraphics[width=0.85\linewidth]{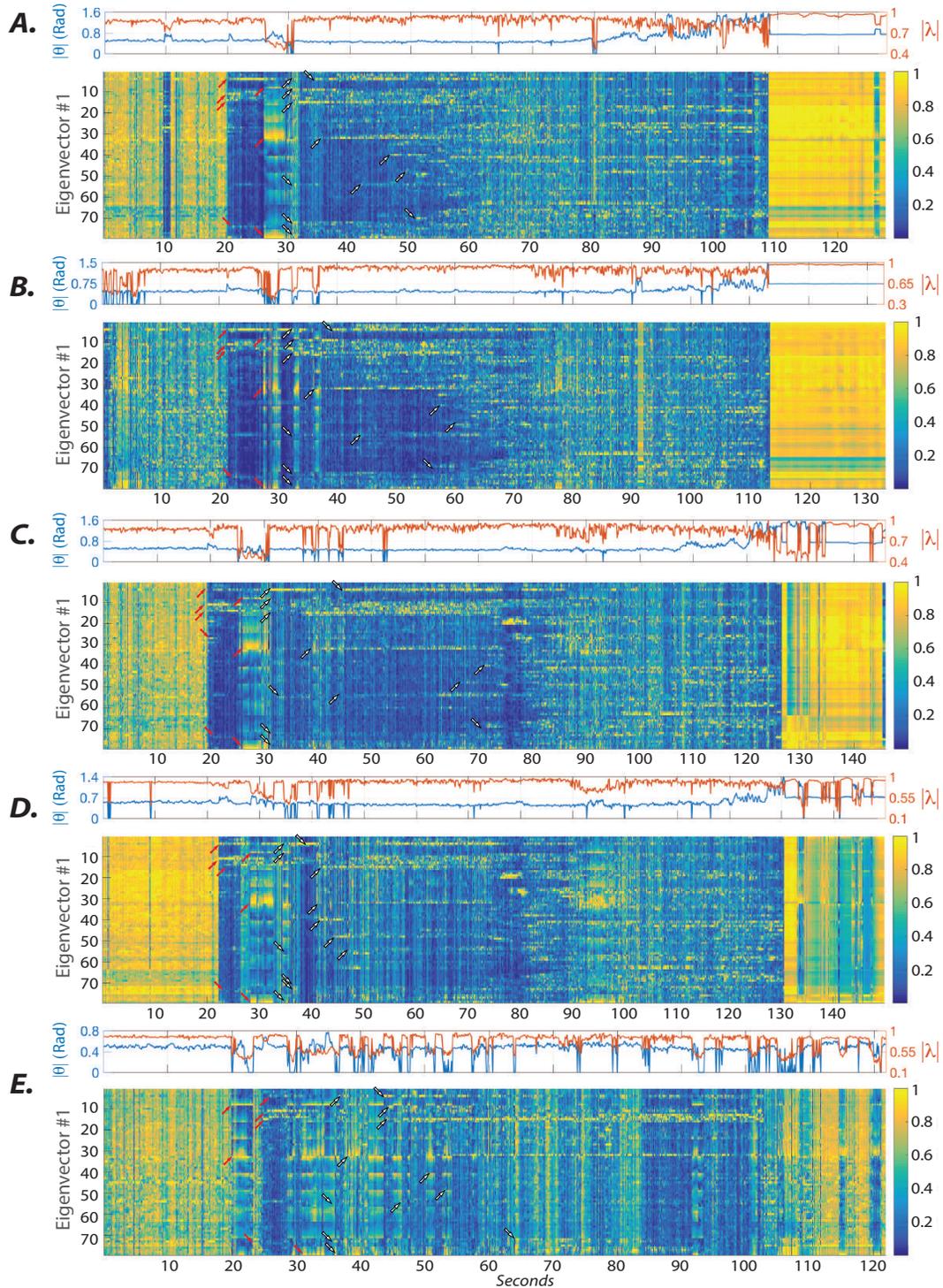}
\caption{\textbf{Evolution of the first eigenmode for subject 3 over several ictal periods, indicating cross-seizure consistency.}  \emph{\textbf{(A)}} The temporal evolution of eigenmode 1 for the seizure previously presented in Fig.~\ref{fig:Figure1}. \emph{\textbf{(B-E)}} The temporal evolution of eigenmode 1 across different seizures from the same subject. We consistently observe periods of approximate time-invariance of eigenvector loading patterns immediately following the onset of the seizure. The seizure shown in panel \emph{\textbf{(E)}} is a complex partial seizure without secondary generalization; here we observe that the eigenmode properties during seizure propagation differ from that observed in seizures that secondarily generalize (panels \emph{\textbf{(A-D)}}). Pointer arrows in all panels are offered as guides to the eye for sensors with high loadings onto the highest frequency eigenvector.} 
\label{fig:Figure2}
\end{figure}

Importantly, the intra-subject similarity of the eigenvectors' temporal evolution at seizure-onset suggests that seizures tend to be led by a small set of cortical regions, consistent with the clinically-accepted notion that seizures originate from specific anatomical \emph{loci}. These general trends were consistently observed across seizures within a single subject (Fig.~\ref{fig:Figure2}, compare panels A-D drawn from four seizures experienced by the same subject). Notably, we observed that while the same qualitative phase-transition at seizure-onset occurs across seizures, the duration of the periods characterized by approximate time-invariance of eigenvector loading patterns can differ across seizures.

Following these qualitative observations, we next performed a quantitative assessment to statistically characterize eigenvector evolution and to support our claim that -- at seizure onset -- the high frequency eigenvectors display approximate time-invariance of loading patterns. Specifically, we calculated the standard deviation of the average pairwise differences between eigenvector loading patterns in neighboring time points (see Methods). We consistently observed periods characterized by greater spatiotemporal invariance than expected in inter-ictal data (Supplementary Materials Fig.~\ref{fig:New_SI_Fig_C}). To show this, we extracted inter-ictal segments ($N= 27.76 \pm 7.35$ per patient) from ECoG recordings that were taken more than $3$ hours before or $3$ hours after a seizure (see Methods). Compared to inter-ictal segments, we observed that ictal onset is marked by a statistically significant increase in spatio-temporal invariance across several eigenvectors with frequency content in the range 4 to 55 Hz (see Supplementary Results for more details). These comparisons between ictal and inter-ictal dynamics demonstrate that the temporal evolution of the eigenvectors reliably highlights the beginning of the seizure-onset period. These findings are consistent with those from the single-subject analysis discussed previously.

\subsection*{Dynamical stability analysis tracks seizure progression and marks seizure offset}

Our results thus far have demonstrated that the seizure-onset period is characterized by approximate time-invariance of eigenvector loading patterns, and that these dynamics are unlike that observed in inter-ictal periods. It is next natural to examine the dynamics of seizure propagation and termination. In both simple-partial and complex-partial seizures, we find that only a few electrodes have high loadings on high frequency eigenvectors, and that these loadings are maintained through the initial time-invariant eigenvector period (Fig.~\ref{fig:Figure3}). Interestingly, these regionally-specific, high eigenvector loadings are greater in magnitude than expected in either inter-ictal or peri-ictal data (see Supplementary Materials Fig.~\ref{fig:New_SI_Fig_H} and Fig.~\ref{fig:New_SI_Fig_I}, respectively). By contrast, the gradual emergence of a large number of electrodes across both hemispheres with high contribution to the eigenvectors marks seizure generalization (Fig.~\ref{fig:Figure3}). 

Seizure offset is marked by sudden changes in the neurophysiological processes that manifest as a return to inter-ictal dynamics. Dynamical stability analysis can characterize these changes at seizure termination through transient changes in the stability metric $|\lambda|$ (see red trace in northern insets of panels Fig.~\ref{fig:Figure1}B-E, Fig.~\ref{fig:Figure2}A-E, and Fig.~\ref{fig:Figure3}B-E). We observed that in all of the CP and CP+GTC samples (after removing one CP+GTC sample with high artifact) the stability of several eigenmodes drops immediately following clinically-marked seizure termination (Fig.~\ref{fig:Offset_Stablity_Drop}). In fact, the average stability across all eigenmodes clearly captures this sudden decrease of the eigenmodes' stability time-locked to seizure offset (Fig.~\ref{fig:Offset_Stablity_Drop}A).  Although the largest decrease in the stability is observed in the CP+GTC samples, normalized average stability curves highlight the similarity between the profile of the stability drop across samples (Fig.~\ref{fig:Offset_Stablity_Drop}B). We quantified the changes in the eigenmodes' stability by calculating the difference between in the average stability over a 3 s window before and after seizure offset. As seen in Fig.~\ref{fig:Offset_Stablity_Drop}C, all the CP and CP+GTC samples show lower stability following seizure offset. Moreover we observed that this decrease in stability is not unique to a single frequency band, but instead can observed across several frequency bands for the majority of samples. Nevertheless, the most robust results were observed for eigenmodes in the $\beta$-band (Fig.~\ref{fig:Offset_Stablity_Drop}C). It is worth noting that this observation did not hold for the few SP seizures, where samples did not reveal a notable pattern of stability fluctuations at the seizure offset. Following seizure termination, the angles and radii returned to inter-ictal values (less stability) quickly, and the eigenvector pattern becomes less time-invariant, and more unstructured (Fig.~\ref{fig:Figure1}B). These observations reveal that the long-lasting and stable epileptic activity ends abruptly at the seizure offset, which is quickly followed by a transition into a regime of relatively fast-damping dynamics.

\subsection*{Inter-subject variability and seizure-type specificity}

\begin{figure}[p] 
\centering
\includegraphics[width=0.85\linewidth]{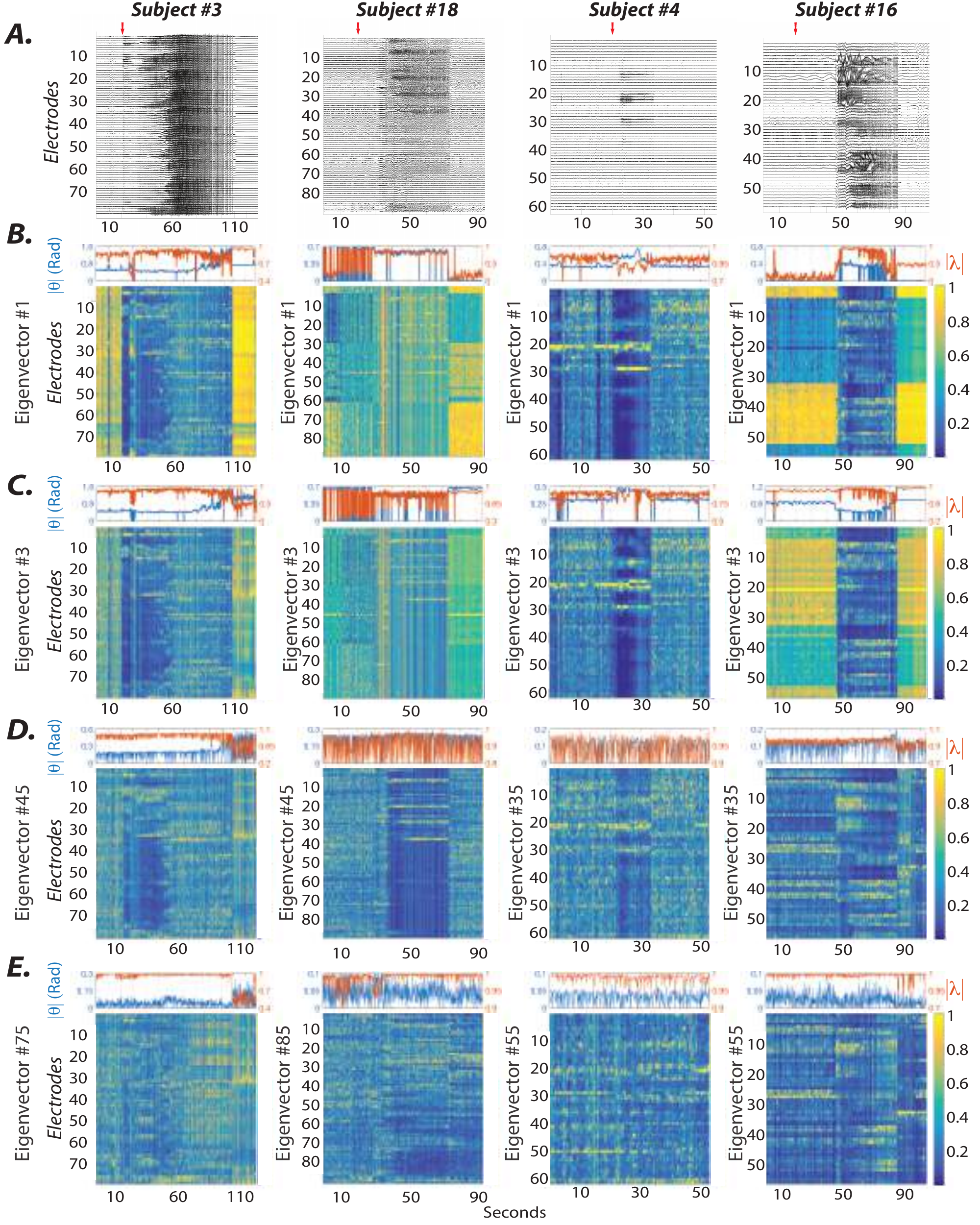}
\caption{\textbf{Evolution of sample eigenmodes of four subjects over ictal periods, indicating inter-subject variability and seizure-type specificity.} \emph{\textbf{(A)}} ECoG signals from all electrodes in four subjects over a single ictal period, from 20 s before seizure onset to 20 s after seizure termination. \emph{\textbf{(B-E)}} The frequency (cyan trace) and stability (red trace) of eigenvalues associated with four representative eigenvectors -- two associated with high frequencies and two associated with low-frequencies -- whose evolution is displayed in heatmaps. The seizures presented for subjects 3 and 16 are complex partial with secondary generalization (compare to Fig.~\ref{fig:Figure2}A-D), while the seizure for subject 18 is complex partial without secondary generalization (compare to Fig.~\ref{fig:Figure2}E), and the seizure for subject 4 is simple partial.} 
\label{fig:Figure3}
\end{figure}

At the group-level, we observe (i) relative time-invariance of eigenvector loading patterns during the ictal period, (ii) this relative time-invariance is unlike that observed in the inter-ictal phase, and is characteristic of eigenmodes representing dynamics in the 4-55 Hz range, and (iii) seizure-onset is characterized by regional localization of eigenvector loading patterns, suggesting the presence of anatomically localized epileptic \emph{foci}. In addition to these group-level findings, we also observed interesting idiosyncrasies across subjects. Specifically, we observed that the relative time-invariance of loading patterns was better captured in the eigenvectors associated with lower frequencies (\emph{i.e.}, 0-8 Hz) for patients with $\delta$--$\theta$ onset epileptic activity as specified in the clinical annotations. Moreover, we observed that ictal samples varied in dynamical stability and in the patterns of eigenvector evolution, suggesting pronounced variability in the spatiotemporal processes corresponding to the evolution and termination of seizures. Finally, we also observed individual differences in the speed of the transition from seizure termination to inter-ictal dynamics: some subjects showed a rapid transition (e.g., see subject 4 in Fig.~\ref{fig:Figure3}) while others showed a more gradual transition (e.g., see subject 3 in Fig.~\ref{fig:Figure3}). These findings point to individual variation in recovery mechanisms that drive brain dynamics from unstable seizure states to normative and stable states.

Importantly, we note that some of the inter-subject variability in the spatiotemporal evolution of the eigenvectors can be explained by seizure type or characteristic frequency. Specifically, we observed that generalized seizures displayed a gradual increase in the number and diversity of regions with high loadings onto the eigenvectors with the largest eigenvalues. In contrast, the contributions of different regions remained relatively unchanged over time in partial seizures (\emph{e.g.}, see subject 4 in  Fig.~\ref{fig:Figure3}). Further, we observed that the estimated angle and stability of eigenmodes during the ictal and inter-ictal periods differed in spatiotemporal frequency, as might be expected due to the different anatomical locations of ECoG sensors. For example, the angle of the highest frequency eigenmode (group average 60.4 $\pm$ 43.9) associated with the pre-ictal period in Fig.~\ref{fig:Figure3} B-C for subjects 3 and 4 is approximately 0.4 rad (i.e., 32.5 $\pm$ 0.9 Hz), whereas subjects 16 and 18 on average displayed much higher ($>$ 0.6  rad) pre-ictal angles (77.2 $\pm$ 35.8 Hz and 51 $\pm$ 4.9 Hz, respectively).

\subsection*{Reliability and replicability across methodological choices}

Finally, since the window length and the local approximation model can affect the estimated dynamics, we tested a variety of window sizes (1, 2, 5, and 10 s) and higher-order linear models to assess the impact of these methodological choices on our results. First, we demonstrate via synthetic examples that the linear model with a short window of 1s is able to accurately capture broad timescales and overlapping dynamics, whereas longer windows lack temporal sensitivity (Fig.~\ref{fig:SI_Figure2}). These numerical studies are consistent with the proposed synthetic example constructed to assess the reliability of the proposed methods (see Supplementary Materials for details). Specifically, longer time windows yield more reliable estimates of low frequency oscillatory modes (as measured by fewer time-points with unstable modes, $|\lambda_i|> 1$), while yielding less reliable estimates of high frequency oscillatory modes (Fig.~\ref{fig:SI_Figure3_new}). We note that the estimated model using empirical data is qualitatively similar for smaller window sizes (\emph{i.e.,} 1 and 2 s windows), and therefore we present the results for the 1 s window. 

Second, to evaluate the impact of the model on our observed results, we considered linear models corresponding to autoregressive models of higher order. In this context, such models are capable of better approximating the local dynamics due to the increased degrees of freedom, which implies that this model will be able to uncover spatiotemporally overlapping dynamics that would not be captured otherwise. In considering the angle and stability of the eigenmodes, our supplementary analysis reveals that the periods of approximate time-invariance in high frequency eigenvectors remains consistently identifiable. In fact, the relatively similar pattern of evolution of the eigenmodes in both autoregressive models suggests that both models capture similar underlying processes during the ictal period (see section entitled ``Dynamical Stability-based Approach using Higher-order Linear Models'' in the Supplementary Materials for more details). These results provide additional evidence supporting the reliability of the proposed criterion in characterizing the beginning and end of the seizure period.

\section*{Discussion}

Here, we seek to understand the spatiotemporal evolution of seizure dynamics using dynamical stability features. The proposed approach decomposes local dynamics into eigenmodes: eigenpairs in which the eigenvectors track the contribution of brain regions to the evolution of the modeled dynamical process and in which the eigenvalues monitor that process's time-scale, frequency, and stability. We observe a relative time-invariance of eigenvector loading patterns during the ictal period. Interestingly, this relative time-invariance in the recruited spatiotemporal processes is unlike that observed in the inter-ictal phase, and is characteristic of eigenmodes representing dynamics in the 4-55 Hz range.  We also find that seizure initiation is characterized by regional localization of eigenvector loading patterns, suggesting the presence of anatomically localized epileptic \emph{foci}. Finally, we observe that seizure termination is marked by a decrease in stability. Then, the angles and radii return to inter-ictal values (less stability) quickly, and the eigenvector pattern becomes less time-invariant, and more unstructured. Our approach directly addresses the challenge that epileptiform activity manifests across a range of temporal and spatial scales, and displays marked inter-sensor dependencies. By assessing interactions between different channels across different time-scales, our approach provides putative mechanisms of seizure generation, progression and termination. More generally, our approach may inform the development of new control strategies for seizure treatment.

\subsection*{Dynamic stability characterization of seizure-onset}

A longstanding theory regarding the dynamical progression of seizures is that they evolve through different stages or dynamical states \citep{Wulsin2013, Rummel2013, Burns2014, Khambhati2015}. While previous studies have motivated approaches to characterize dynamical and topological features of these states, a systems-level understanding of how these dynamical transitions occur has been elusive. Theoretical and experimental evidence points to a mechanism of seizure generation and progression in which the epileptic brain is more susceptible to transitions between stable and asymptotically stable dynamical states. However these studies undersample the spatiotemporal statistics of the network by relying on information acquired only from a single channel \citep{jirsa2014nature, Worrell2002Evidence}. 

In contrast, here we develop and apply a novel approach that enables us to characterize the onset and termination of seizures in terms of the dynamical modes of the neural system by accounting for the epileptic network's rich spatiotemporal structure. First, we find that the dynamics immediately following the EEC demonstrate an approximate spatio-temporal invariance of eigenvectors associated with several frequency bands (ranging from 4-55 Hz), along with stereotypical shifts in the frequency and stability of eigenmodes common across patients in peri-ictal events. These observations suggest that the emergence of persistent focal oscillatory sources eventually drives the underlying system into an ictal regime \citep{perucca2014intracranial,curtis2016gabaergic, Khambhati2017Recurring}. Second, we find that seizure termination is characterized by a transition from a stable state in which a few regions have high eigenvector loadings, to a more asymptotically stable state in which many regions have comparable eigenvector loadings, consistent with the dynamics observed during inter-ictal periods. 

Importantly, eigenmodes of the system distinguish between commonly diagnosed seizure types. In line with current clinical classification of seizure types \citep{fisher2017operational}, we found that partial seizures -- whose pathological dynamics remain focal -- present approximate time-invariance of several eigenvectors during the entire seizure with contributions from only a few regions. In contrast, generalized seizures -- whose pathological dynamics progress to bilateral tonic--clonic -- present a heterogenous distribution of patterns as seizures spread to different regions and eventually across the hemispheres. Importantly, the observation that different sorts of seizures have different patterns of inter-sensor dynamics is not wholly new and has also been reported in studies of time-resolved patterns of functional connectivity in ECoG \citep{khambhati2016virtual,schindler2006assessing}. Indeed, it is interesting to speculate that the previously reported seizure-type-specificity of the synchronizability of the ECoG network estimated from functional connectivity may be related to differences in the underlying process, as revealed by our modeling approach. More generally, a fundamental understanding of the processes driving different seizure types could inform the development of seizure-type-specific interventions in the form of stimulation via implantable devices perhaps through objective functions that seek to dynamically control parameters such as frequency and stability of the neural system.

\subsection*{Underlying drivers of seizure-onset and prevention}

Our methodological approach enables an assessment of changes in dynamical stability between non-seizure and seizure periods. We find that the inter-ictal period is characterized by high temporal variance of the system's eigenvectors, reminiscent of findings in prior work suggesting that the epileptic network traverses different dynamical states more frequently during the inter-ictal period than during the ictal period \citep{Khambhati2015, Khambhati2017Recurring, Burns2014}. We also find that the inter-ictal period is characterized by limited changes in frequency and stability. These insights from the dynamical stability-based characterization suggest that a possible mechanism to mitigate seizure propagation would be to stimulate different regions towards a homogenization of the contributions of these regions to the local dynamics. That is, rather than only targeting seizure-onset nodes, it may be important to stimulate nodes in the surrounding tissue to create heterogeneity in the dynamics, thereby increasing resilience to changes in the stability of the dynamical process. A similar suggestion has been made previously: that degree heterogeneity and lower synchronizability of the network relates to increased focalization of seizures, and that the largest contributors to this relationship are found in the surrounding tissue, rather than in the seizure onset zone \citep{khambhati2016virtual,weiss2013ictal,schevon2012evidence}. This study predicted that if these key electrodes in the surrounding tissue were removed or virtually lesioned, one would expect a profound perturbation of the entire network's synchronizability. An important area for future work is to understand the degree to which stimulation paradigms that alter synchronizability also alter dynamical stability, as studied here.

Importantly, prior work has outlined the general aim of stimulating a region such that the surrounding tissue changes its properties, while ensuring that the stimulation protocol itself satisfies certain constraints \citep{alonso2015influence}. Example constraints include not targeting regions that are deemed unsafe or unethical to stimulate, and not using stimulation parameter values (such as for frequency and magnitude) that are deemed ineffective, not technologically feasible, or unsafe for human tissue. In principle, our stability-based assessment could inform similar studies, by defining a time-varying stimulation strategy that alters the dynamical stability in such a way as to quiet or prevent seizures while minimizing the potential of pushing the system into a dynamical stability regime that is uncontrollable. Indeed, theoretical work based on similar principles to those we use here to model the ECoG data has begun to develop a framework to specify such a stimulation strategy \citep{pequito2017spectral}. In principle, should this theory prove effective, it could be used to inform ECoG sensor placement during installation of implantable devices, and sensor stimulation parameters by closed loop measurement and manipulation of system stability. The key advantage of this framework is that -- if validated clinically -- it could offer a thoroughly quantitative approach to stimulation protocols for implanted devices.

More speculatively, the features that we obtain from eigenmode decomposition could be used as quantifiable performance objectives for optimizing stimulation parameters in closed-loop implantable devices \citep{stacey2008technology,6428623}. Specifically, this approach would depart from the current practice of manual, intermittent, off-line tuning of deep brain stimulation parameters such as amplitude and frequency based on the count of seizures experienced by the patient. Instead, the features that characterize the seizure-onsets using the dynamical stability-based approach can be used to select channels for feature extraction in a closed-loop assessment and stimulation paradigm \citep{wu1998segmentation}. For example, sensors included in the calculation of time-resolved seizure probability could be limited to those for which eigenvalue-eigenvector features provide reliable estimates of stability. This data reduction could simplify simplify and speed calculations, while maintaining optimal predictions about effective stimulation protocols.

\subsection*{Methodological Considerations}

Several methodological considerations are pertinent to this work. First, the proposed analysis does not directly allow for the assessment of underlying biological mechanisms. Second, although it is likely that the underlying dynamical process captured by ECoG data is non-linear, the dynamical \emph{stability}-based approach allows for the analysis of global nonstationarity and non-linear dynamics by fitting a linear model to short temporal windows. Third, to assess reliability and reproducibility, we examined the robustness of our results to different temporal windows and different models (see Supplement). Our reliable findings therefore provide evidence for a more general phenomenon that cannot be explained by either noise or non-linearity ~\citep{netoff2004analytical}. Fourth, we have demonstrated that the dynamical stability analysis allows us to identify similar changes in the underlying process during the seizure onset period for different samples across patients. Despite our rather homogeneous dataset, which includes a large number of seizures with overlapping features, heterogeneity is still present in the evolution of seizures across subjects and samples. It would be interesting in future to examine changes by brain region, assessing for example whether a frontal onset pattern differs significantly from a temporal onset pattern, or whether a right-dominant onset pattern differs significantly from a left dominant onset pattern. It would also be interesting to examine the duration of the reduced stability during the post-ictal period, and determine whether that duration differs by seizure type. Many factors such as the location and number of electrodes, sampling rate, recording noise, and type of seizures partially explain the observed inter-subject variations. Nevertheless, the dynamical stability analysis provides an avenue for characterization of individual patients' idiosyncrasies and for establishing individualized features that can be used as biomarkers.

\subsection*{Future Directions}

In summary, we provide a dynamical stability-based characterization of the onset of seizure activity that hinges on an eigenmode decomposition of a computationally tractable model. Furthermore, the proposed approach uses linear stability analysis to reveal that seizures are a stereotyped behavior of brain networks that can be leveraged to aid the classification and identification of seizures. This model is an important candidate for use in sensing and actuation design, as well as for new stimulation strategies to be integrated into a future generation of implantable devices. The model provides fine-scale temporal assessment of seizure dynamics, and can be extended to model the impact of control strategies on seizure evolution, with the goal of seizure prevention.

\section*{Acknowledgements}
     
D.S.B., A.A., and A.N.K. would like to acknowledge support from the John D. and Catherine T. MacArthur Foundation, the Alfred P. Sloan Foundation, the Army Research Laboratory and the Army Research Office through contract numbers W911NF-10-2-0022 and W911NF-14-1-0679, the National Institute of Health (2-R01-DC-009209-11, 1R01HD086888-01, R01-MH107235, R01-MH107703, R01MH109520, 1R01NS099348, R21-M MH-106799, the Office of Naval Research, and the National Science Foundation (BCS-1441502, CAREER PHY-1554488, BCS-1631550, and CNS-1626008). K.D. acknolwedges support from K23NS073801-01, and UH2-NS095495-01. S.P. and G.J.P. are supported in part by the TerraSwarm Research Center, one of six centers supported by the STARnet phase of the Focus Center Research Program (FCRP) a Semiconductor Research Corporation program sponsored by MARCO and DARPA. The content is solely the responsibility of the authors and does not necessarily represent the official views of any of the funding agencies.

\clearpage 
 \newpage

\bibliographystyle{plain}
\bibliography{journalBiblio,selfreg}

\clearpage
\newpage

\begin{center}
\LARGE{Supplementary Materials\\}
\Large{for\\``Parsing spatiotemporal dynamical stability in ECoG during seizure onset, propagation, and termination''}
\end{center}
~\\
~\\
\section*{Supplementary Methods}

In this supplementary methods section, we provide complementary information that aims to sharpen the intuition behind the dynamical stability-based approach exercised in the main text. First, we present an application of the dynamical stability-based approach to synthetic data. Second, we show that high-order linear models do not qualitatively change the proposed seizure-onset characterization.

\subsection*{Synthetic example}

\begin{figure}
\centering
\includegraphics[width=0.8\linewidth]{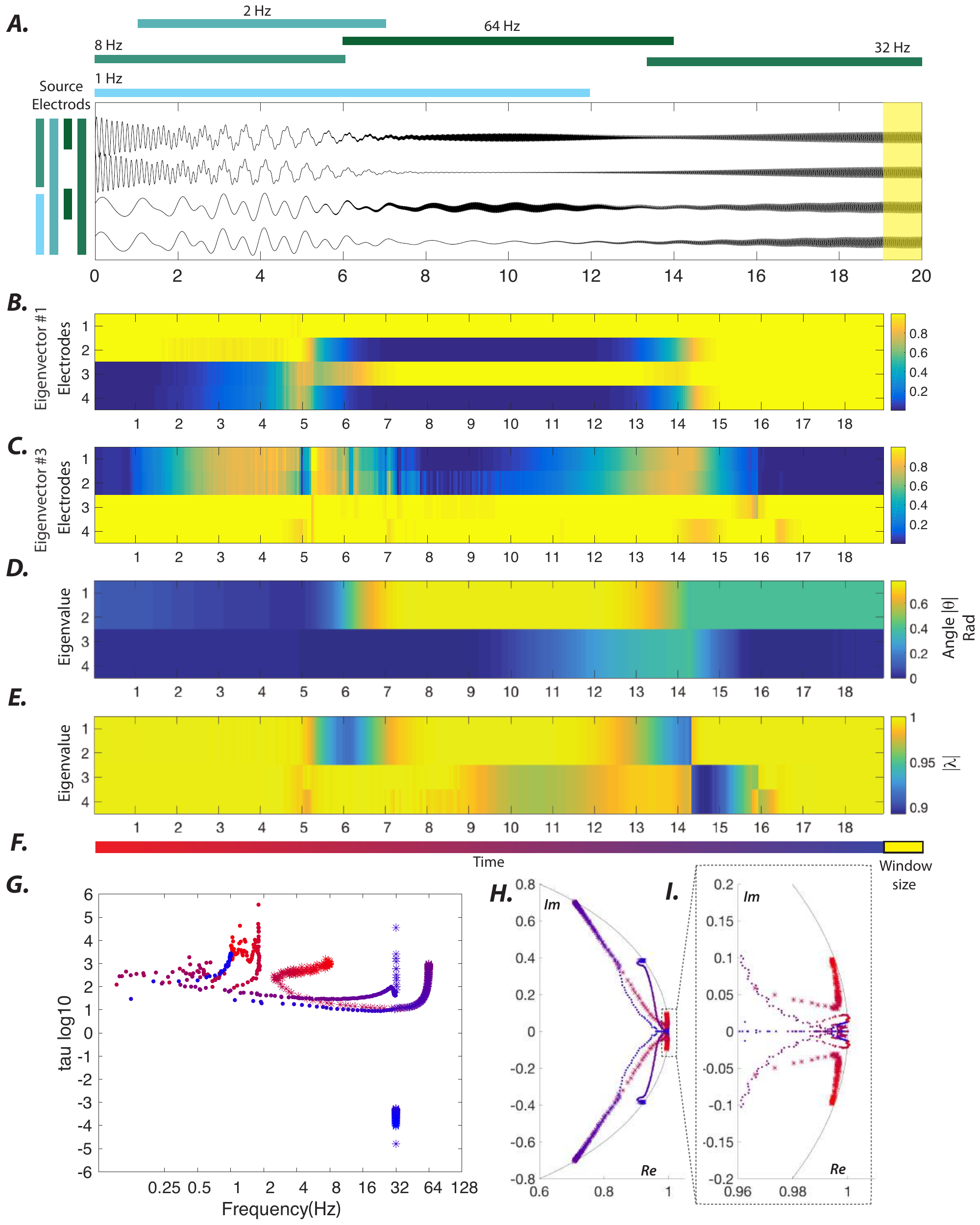}
\caption{\textbf{Dynamical stability-based characterization of synthetic time-series}. \emph{\textbf{(A)}} We constructed surrogate data composed of time-series that capture an underlying process: an enveloped (exponential, Gaussian) sinusoid. Briefly, the simulated time-series capture various partially-overlapping spatiotemporal patterns of oscillations with varying frequencies. The colored bar at the top of the panel represents the duration over which the oscillation with that specific frequency is present. We used a 2-STD interval for oscillations with a Gaussian amplitude profile, and we used $1-100 \%$ maximum amplitude of the signals with exponential amplitude profiles. The colored bars on the left highlight the sources of the oscillations, where the top source leads the following source by $\pi/2$. The yellow shaded area on the right highlights the size on the moving window. \emph{\textbf{(B-C)}} The absolute values of the normalized eigenvectors 1 and 3 obtained using a window of 1 s duration moved by 100 ms shifts. \emph{\textbf{(D-E)}} The evolution of the angle and radius associated with the four eigenvectors of the 4-sensor system. \emph{\textbf{(F)}} The temporal progression of the sliding window illustrated in changing hues of red; the yellow box represents the relative size of the moving window used to estimate the linear model. \emph{\textbf{(G)}} The temporal evolution of the frequency (Hz) and damping ($log10( \tau)$) associated with the first eigenmode (`dot') and third eigenmode (`star'). \emph{\textbf{(H)}} The temporal evolution of the four eigenvalues in the Argand complex plane. \emph{\textbf{(I)}} A zoomed-in version of panel \emph{\textbf{(H)}} to show the changes in the angle and radius of the low angle eigenvalues in greater detail. Panels \emph{\textbf{(G-I)}} use the same color-coding as in panel \emph{\textbf{(F)}}. } 
\label{fig:SI_Figure1}
\end{figure}

Here we provide a pedagogical example of the dynamical stability-based approach, and demonstrate its ability to accurately uncover a planted locus, frequency, and damping of oscillatory dynamics. The synthetic dataset that we study consists of four channels associated with four oscillatory sources, sampled at 512 Hz. Each source is modeled as an enveloped (exponential, Gaussian) sinusoid, and the four sources have a fixed lag between them to induce directionality of information flow across sensors. Specifically, the baseline activity of each channel is modeled as follows:
\[
y=sin(ax+b) e^{cx}
\]
where $a$ determines the frequency of the oscillation, $b$ captures the phase lag between electrodes, and $c$ determines the damping of the oscillations. Negative values of $c$ were used for early damping oscillations while positive values of $c$ were used for late emerging high frequency oscillations. In addition to this baseline activity, we also modeled transient activity that emerges and disappears in brief intervals during the recordings. We model these transient dynamics as follows:
\[
y= sin(ax+b)\frac{1}{\sqrt{2\sigma^{2} \pi}}  e^-{\frac{ (x-\mu)^{2}}{2\sigma^{2}}}
\]
\noindent where $\mu$ is the mean and $\sigma$ the standard deviation of a normal distribution. Finally, different simulated patterns of oscillatory activities over time were linearly superimposed to create the activity of individual channels. Additional details regarding this process of sensor time series creation is depicted schematically in Fig.~\ref{fig:SI_Figure1}A.

After constructing the sensor time series, we next applied the dynamic stability approach to determine its ability to accurately uncover the planted structure of the oscillatory dynamics. First, we show the eigenvector evolution associated with the highest frequency dynamics in the signals (Fig.~\ref{fig:SI_Figure1}B), and then we show the eigenvector evolution associated with dynamics in a lower frequency (Fig.~\ref{fig:SI_Figure1}C). We observe that the locus of higher and lower frequency oscillations at every time point can be identified via electrodes with relatively higher values in the eigenvectors $1$ and $3$, respectively. We also display the stability, $| \lambda |$ associated with those same two eigenvectors (Fig.~\ref{fig:SI_Figure1}D-E). We notice that the changes in the angle match the changes of the underlying frequency of oscillations, while the intervals with radius $\approx 1$ highlight periods of instability. Lastly, we studied the temporal evolution (Fig.~\ref{fig:SI_Figure1}F) of the statistics of interest derived from the eigenmode decomposition (Fig.~\ref{fig:SI_Figure1}G-I). In general, we observe that the dynamic stability approach is able to accurately uncover the planted structure of the oscillatory dynamics, both in terms of their frequency, and in terms of their stability, as a function of time.

\begin{figure}
\centering
\includegraphics[width=0.85\linewidth]{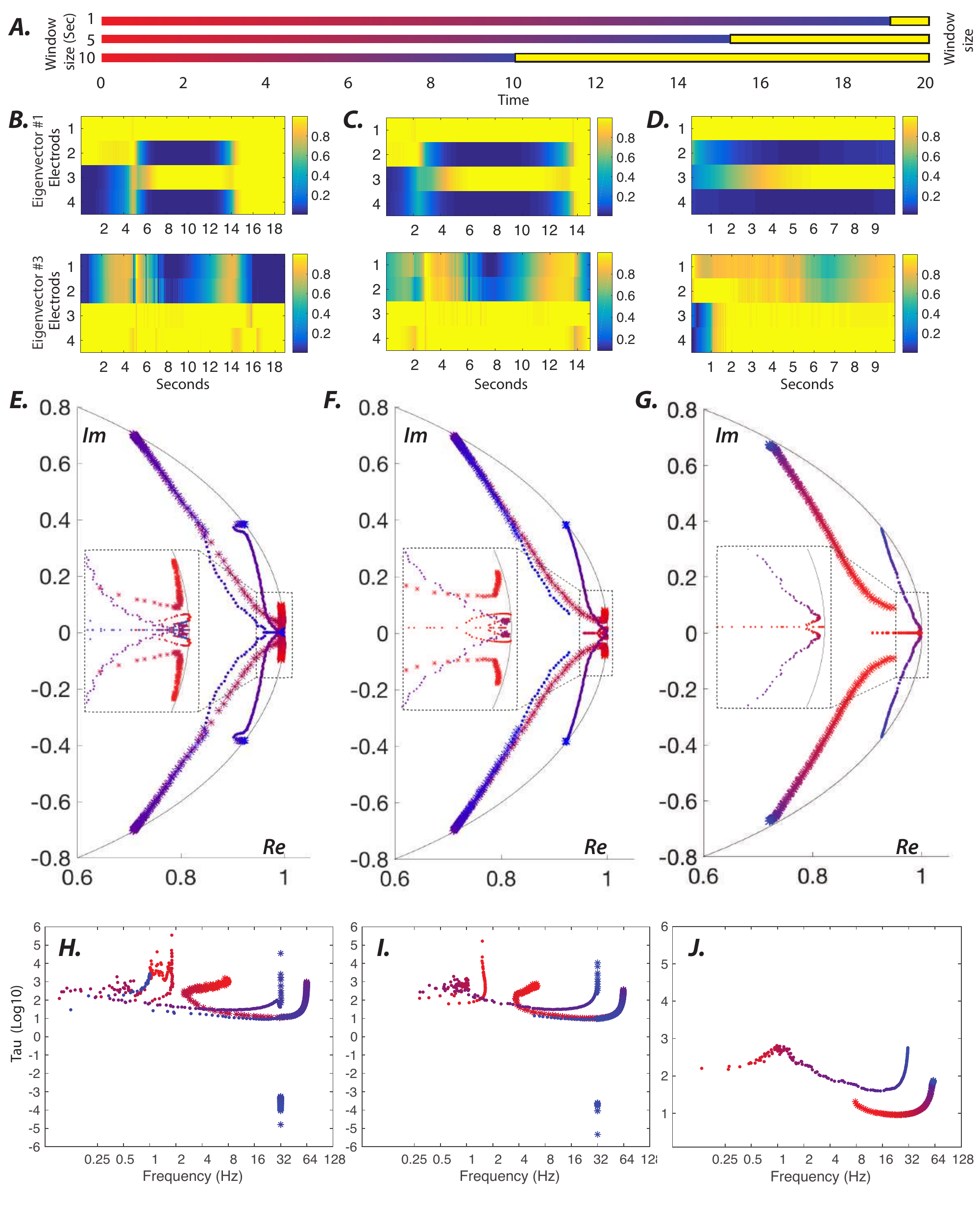}
\caption{\textbf{Effect of window size on the dynamical stability-based characterization of synthetic time-series.}  \emph{\textbf{(A)}} Different sliding windows for the synthetic data shown in Fig.~\ref{fig:SI_Figure1}A are encoded by the yellow box, and the corresponding time progression is color-coded as in Fig.~\ref{fig:SI_Figure1}F. \emph{\textbf{(B-D)}} The absolute values of the normalized eigenvectors 1 and 3 obtained using a window  of 1, 5, and 10 s duration moved by 100 ms shifts \emph{\textbf{(E-G)}} The Argand diagram of the temporal evolution of the four eigenvalues for 1, 5, and 10 s windows, respectively. Insets: Zoomed-in version of the same plots to show the changes in the angle and radius of the low angle eigenvalues in greater detail. \emph{\textbf{(H-J)}} The temporal evolution of the frequency (Hz) and damping ($log10( \tau)$) associated with the first eigenmodes (`dots') and third eigenmodes (`stars') associated with 1, 5, and 10 s windows. Panels \emph{\textbf{(E-J)}} use the same color-coding as in panel \emph{\textbf{(A)}}.}
\label{fig:SI_Figure2}
\end{figure}

Lastly, we asked whether the accuracy of the dynamic stability approach depended on the time window over which we estimated the ARMA model. Intuitively we might imagine that if the time window is too large, we would not be able to accurately specify short time-scale changes. We therefore applied the dynamic stability analysis to the surrogate data windowed over a range of temporal lengths (Fig.~\ref{fig:SI_Figure2}). We found that the frequency and stability estimates did indeed depend on the length of the time window over which the ARMA model is fit. Specifically, the estimated frequency and damping of the oscillations were reduced progressively for larger window sizes, likely due to the presence of temporally overlapping patterns of oscillations during the time-windows considered for estimation.

\subsection*{Dynamical stability-based approach using higher-order linear models}

\begin{figure}
\centering
\includegraphics[width=0.85\linewidth]{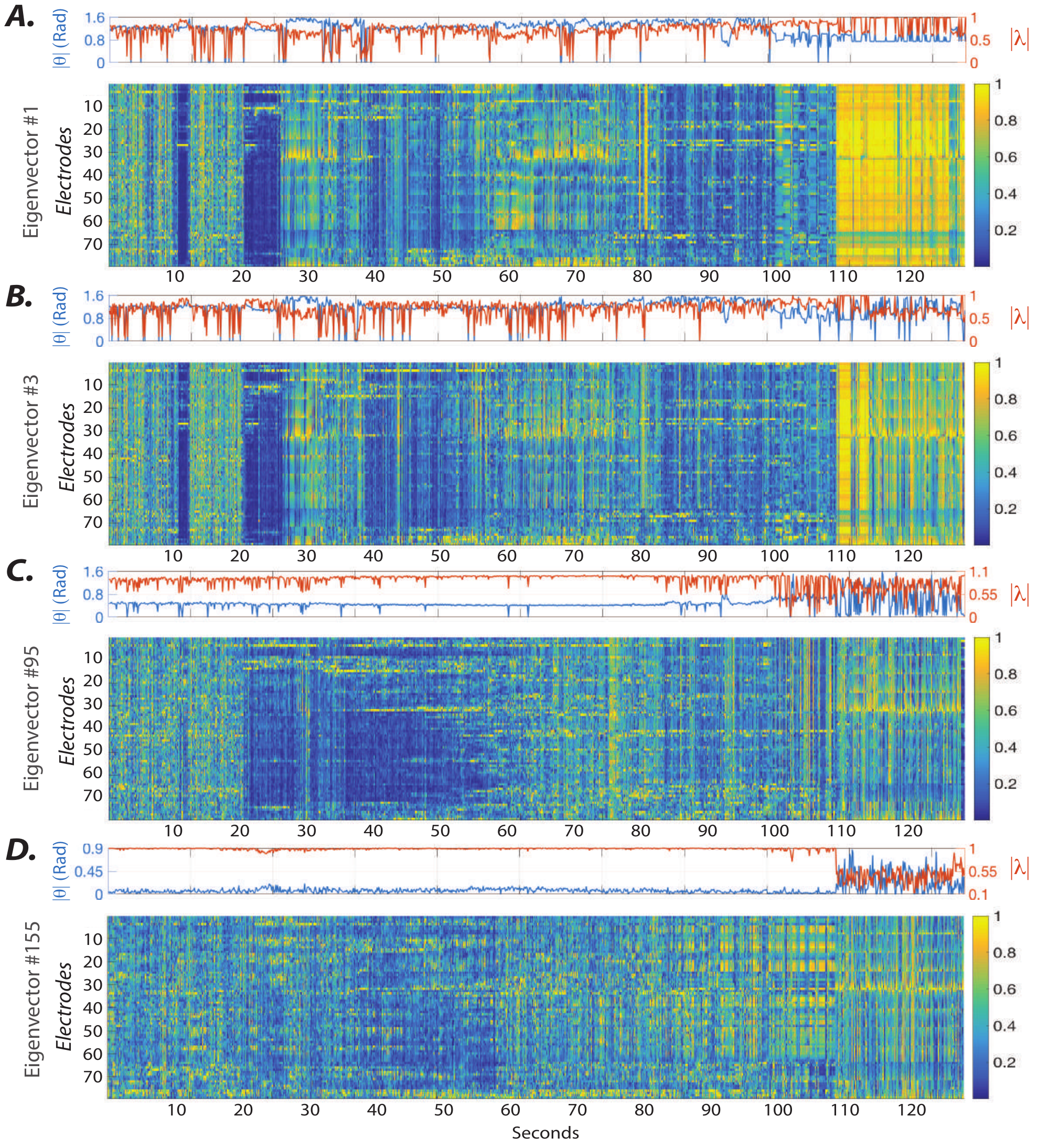}
\caption{\textbf{Evolution of eigenmodes estimated from a second order autoregressive model over the ictal period shown in Fig.~\ref{fig:Figure1}A.}  \emph{\textbf{(A-D)}} The frequency (cyan trace) and stability (red trace) of eigenvalues associated with four representative eigenvectors -- two associated with high frequencies and two associated with low-frequencies -- whose evolution is displayed in heatmaps. Naturally, these models are better able to capture spatiotemporal high frequency patterns. For instance, the period of approximately time-invariant eigenvector loading patterns shown in Fig.~\ref{fig:Figure1}B between $27$-$33$ s is here marked by high frequency transient oscillatory activity present across a large number of electrodes, and extending over a long time period. } 
\label{fig:SI_Figure5}
\end{figure}

The dynamical stability-based approach relies on a first-order approximation of a non-linear process, and therefore only provides a local approximation of the underlying dynamics. Although computationally efficient and simple, it is important to consider the drawbacks of this approach in comparison to approaches that employ higher-order models. Because the data considered is in discrete-time and the linear model used can be understood as a coupled autoregressive model of first-order, we assess how a coupled autoregressive model of higher-order might impact the observed results. Towards this goal, we used the Schwarz's Bayesian criterion \citep{neumaier2001estimation} to estimate the order of the coupled autoregressive model. For a 1 s time-window, the order of the coupled autoregressive model is generally lower than three, across time and subjects. Therefore, we contrast the results obtained with a coupled autoregressive of order two model (Fig.~\ref{fig:SI_Figure5} in this Supplement) against those reported in Fig.~\ref{fig:Figure1} of the main manuscript.

We observe notable differences between the dynamics of the two models, as expected from a theoretical point-of-view: higher-order models have extended memory that enables the monitoring of the spatiotemporal lag in the evolution of the underlying dynamical process. First, the angles of the high frequency eigenmodes of the second-order autoregressive model are much higher than those of the first-order autoregressive model (compare Fig.~\ref{fig:SI_Figure5}A to Fig.~\ref{fig:Figure1}A). Second, the eigenvectors of the first-order autoregressive model suggest that the second period of high frequency dynamics following the EEC onset is only present between $27$-$33$ s, while the eigenvectors of the second-order autoregressive model suggest that these dynamics extend from $27$-$80$ s (compare Fig.~\ref{fig:Figure1}A and Fig.~\ref{fig:SI_Figure5}A). These results support the presence of temporally overlapping high frequency dynamics that cannot be captured by the first-order autoregressive model. Despite these differences, both models present eigenvectors associated with high frequencies that exhibit approximately time-invariant loading patterns at seizure onset and termination. Moreover, they both show the contribution of only a few regions for the dynamic process associated with the spatiotemporal high frequencies at seizure onset, and at seizure termination they present a sudden transition to capture a larger number of contributions with almost equal weight across different neocortical regions. These similarities between the two models corroborate the capability of the proposed dynamical stability-based approach to characterize seizure dynamics.

\newpage

\section*{Supplementary Results}

\subsection*{Seizure offset is marked by a sudden decrease in the stability of several egienmodes}

\begin{figure}
\centering
\includegraphics[width=1\linewidth]{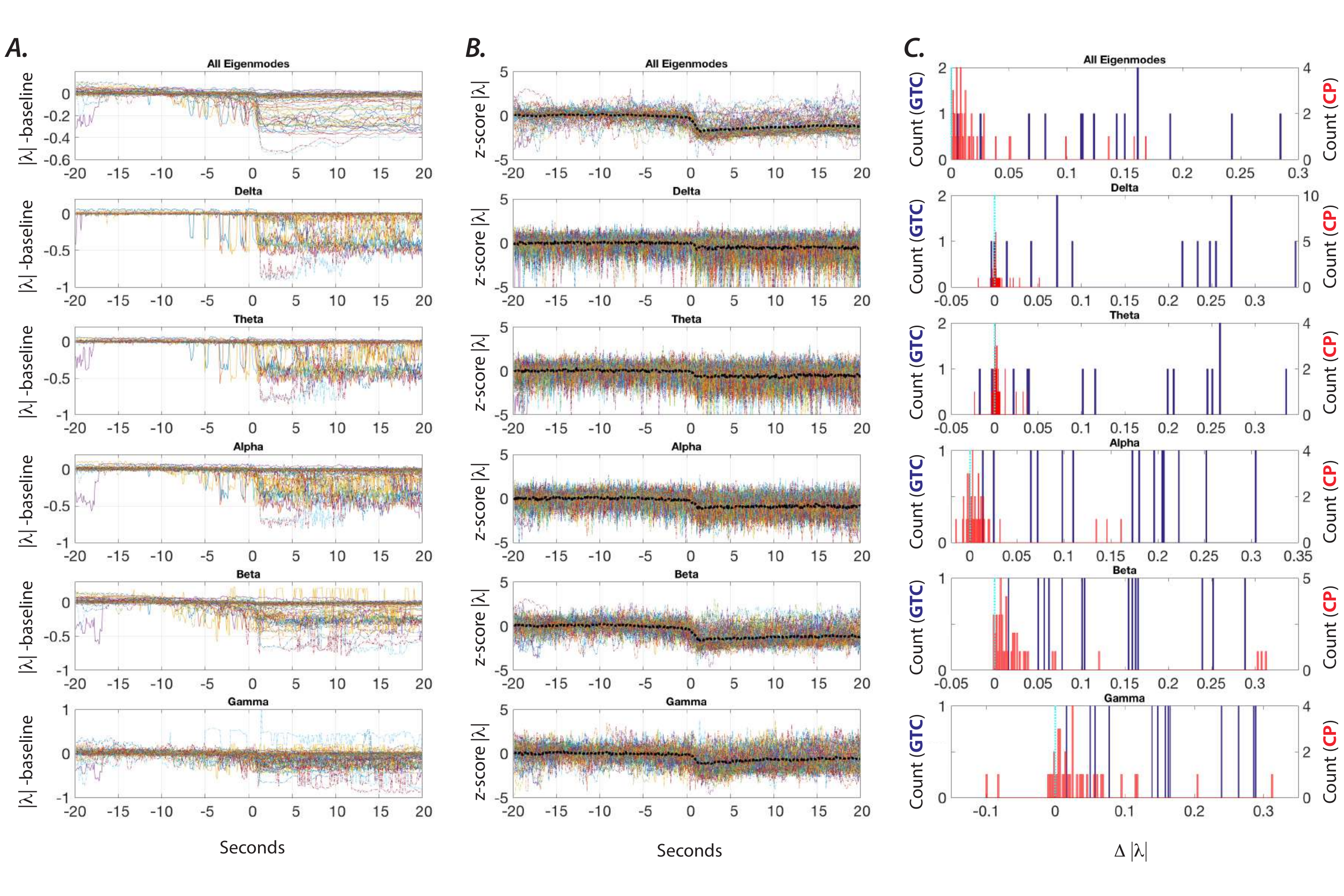}
\caption{\textbf{Seizure offset is marked by a sudden decrease in the stability of several egienmodes.}  \emph{\textbf{(A)}} (Top) The average stability of all eigenmodes over a 40 s window centered around the ictal offset, and plotted for all CP (dashed line) and CP+GTC (solid line) samples, after removing a single CP+GTC sample with high recording noise and all samples from subjects 14,15, and 21 (insufficient clinical annotations). The y-axis shows the average stability values following the subtraction of a 20 s pre-offset baseline. Note that large drops in the stability are present in all CP+GTC samples. The results for the eigenmodes, after band-pass filtering based on the eigenmodes' average frequency are presented from top to bottom to reflect $\delta$, $\theta$, $\alpha$, $\beta$, and $\gamma$.  \emph{\textbf{(B)}} Average stability plots normalized calculating the $z$-score. The dashed black line represents the event-related average of samples. \emph{\textbf{(C)}} Histogram of the difference between the average stability over a 3 s window before and after ictal offset. The blue histogram represents the CP+GTC results and the red histogram  represents the CP results. Note that all samples show higher average pre-offset average stability across all eigenmodes (Top). Although this effect is not unique to an individual frequency band, the most robust results are observed in the $\beta$-band.}
\label{fig:Offset_Stablity_Drop}
\end{figure}

Almost all ictal samples (with the exception of SP) reveal a similar decrease in the stability of several eigenmodes following ictal offset. The average stability of all eigenmodes at ictal offset -- which we plot for all CP and CP+GTC samples (samples from subjects 14,15, and 21 are not presented due to insufficient clinical annotations) in Fig.~\ref{fig:Offset_Stablity_Drop}A (top) -- highlights the sudden change in the stability that is time-locked to the seizure offset. Although the magnitude of the decrease in the stability is much larger in most GTC samples (as seen in Fig.~\ref{fig:Offset_Stablity_Drop}A\&C), CP samples show comparable changes though at smaller magnitude. This similarity is better captured in the normalized ($z$-score) event-related average plots shown in Fig.~\ref{fig:Offset_Stablity_Drop}B. Although the offset drop in the stability is not limited to a single frequency band, calculating the difference between the average stability over a 3 s window before and after ictal offset reveals that all CP and CP+GTC samples (after removing one CP+GTC sample with high recording noise) show a large offset stability drop in the $\beta$-band. These observations suggest that the estimated dynamics immediately following ictal offset are more asymptotically stable -- i.e., the post-ictal oscillations damp much faster than pre-offset (ictal) oscillations.

\subsection*{Impact of the time-window length in the dynamical stability-based approach}

\begin{figure}
\centering
\includegraphics[width=0.85\linewidth]{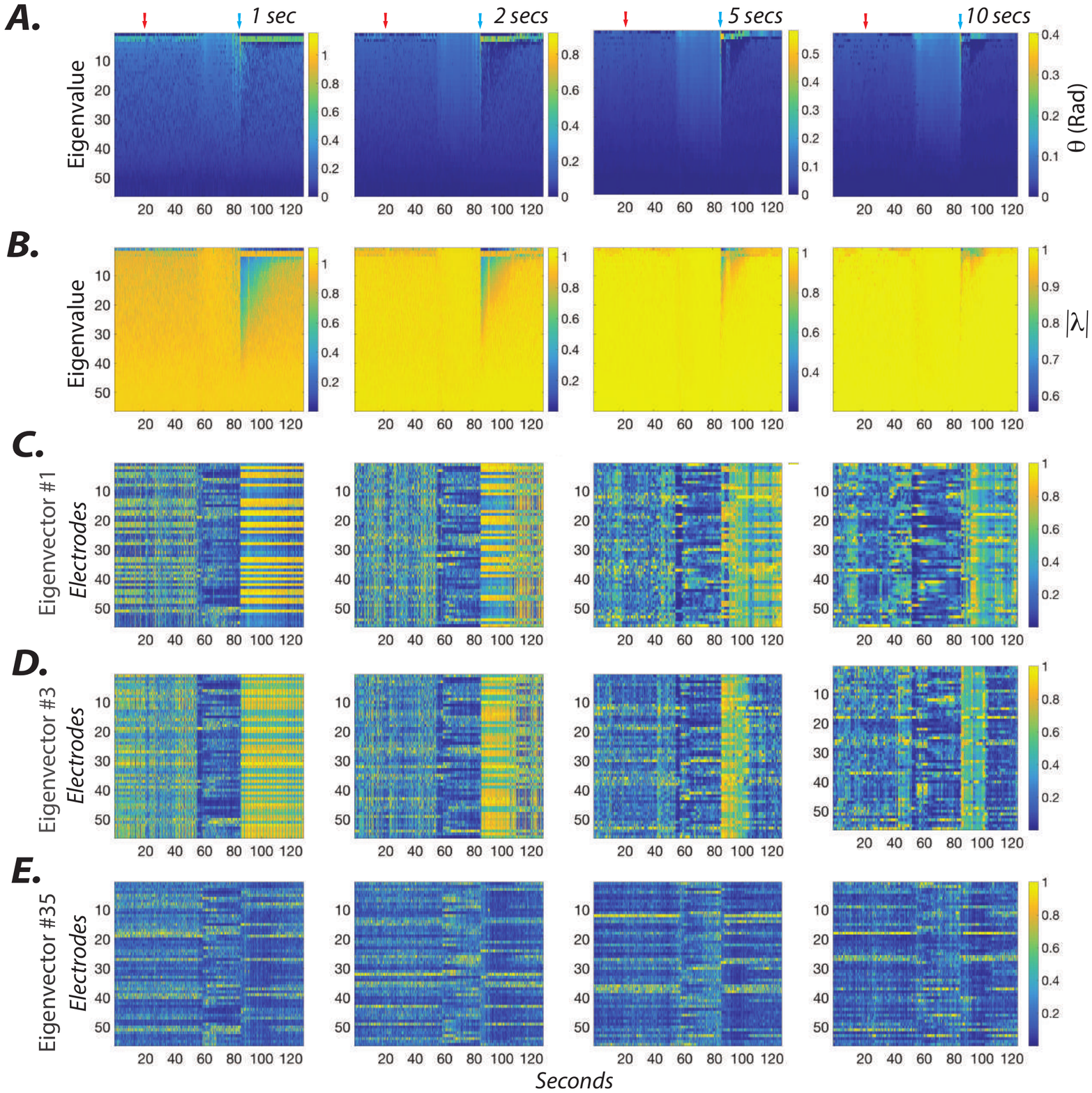}
\caption{\textbf{Effect of window size on the dynamical stability-based characterization of ictal periods.}  \emph{\textbf{(A-B)}} For the example seizure from subject 16 shown in Fig.~\ref{fig:Figure3}A, we show the time evolution of the angles and stability of the eigenvalues associated with the different eigenmodes, for different window sizes (1, 2, 5, and 10 s; shown across different columns). The red and cyan arrows highlight the onset and offset of the ictal period, respectively. \emph{\textbf{(C-E)}} The temporal evolution of three representative eigenvectors -- two associated with high frequencies and one associated with low-frequencies.}
\label{fig:SI_Figure3_new}
\end{figure}

Here we ask whether the results of the dynamic stability approach depend on the time window over which we estimate the ARMA model. Our intuition is that if the time window is too large, we might not be able to accurately specify short time-scale changes. We therefore consider 1, 2, 5 and 10 s time-windows, while maintaining 100 msec shifts. The calculations provide similar intuitions to those we obtained from considering the synthetic data: we observe a loss of sensitivity to high frequency dynamics as the length of the time-window increases (Fig.~\ref{fig:SI_Figure3_new}). Despite these window-dependent effects, our main results still hold: (i) the eigenvectors associated with high frequencies present approximate time-invariance of loading patterns for a relatively long period of time, with the contribution of a few regions at the onset of the seizure, and (ii) there is a sudden change at the end of the seizure where a larger number of regions contribute in an approximately equal manner to the evolution of the spatiotemporal high frequency process. These consistent findings bolster confidence in the reported results.

\subsection*{inter-ictal dynamical stability-based characterization}

\begin{figure}[p] 
\centering
\includegraphics[width=0.85\linewidth]{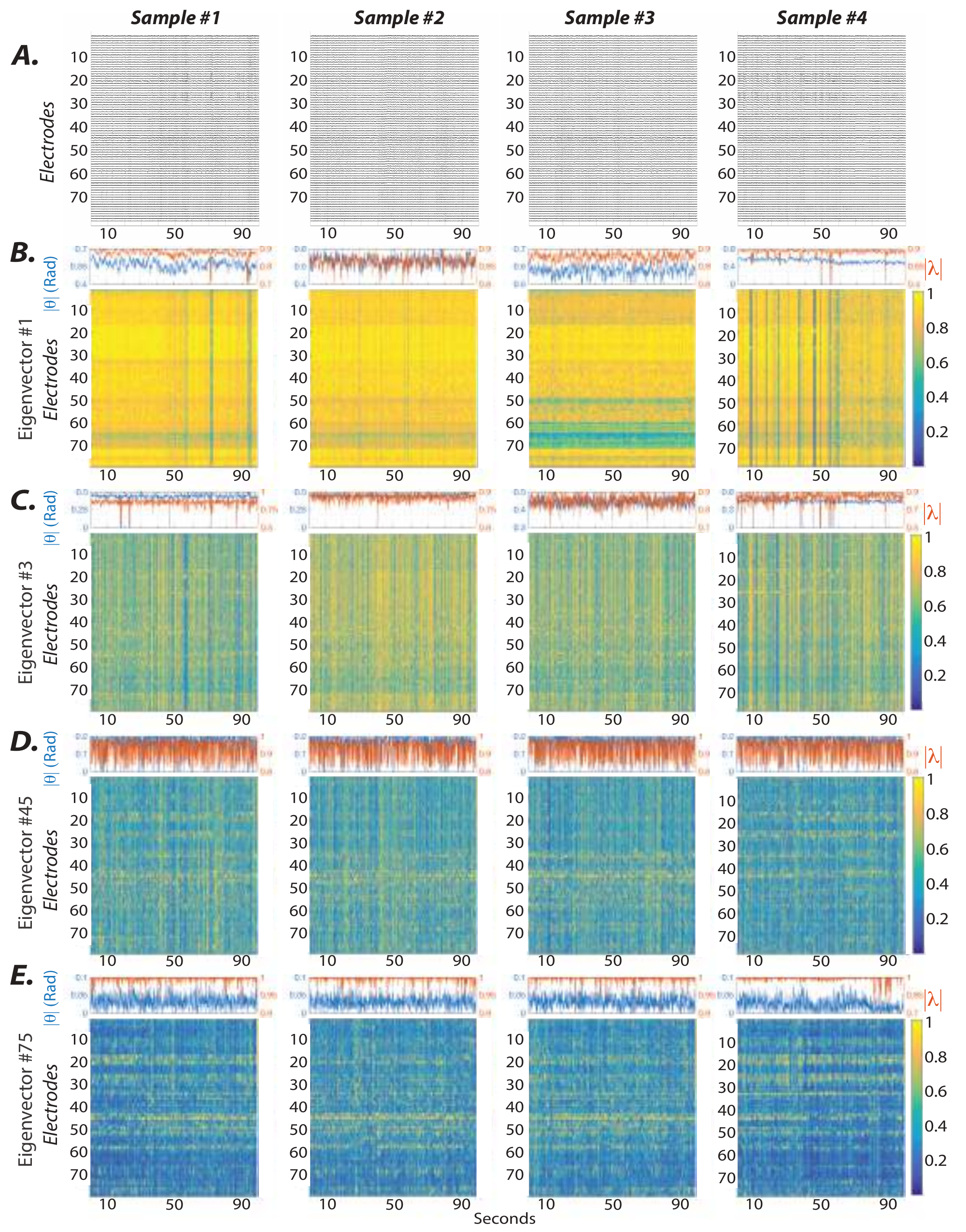}
\caption{\textbf{Evolution of sample eigenmodes for four sample inter-ictal periods of subject 3.} \emph{\textbf{(A)}} ECoG signals from 79 electrodes in subject 3 over four sample ictal periods (across the columns), from 20 s before seizure onset to 20 s after seizure termination. \emph{\textbf{(B-E)}} The frequency (cyan trace) and stability (red trace) of eigenvalues associated with four representative eigenvectors -- two associated with high frequencies and two associated with low-frequencies -- whose evolution is displayed in heatmaps. } 
\label{fig:SI_Figure4}
\end{figure}

While our study focuses on ictal dynamics, it is also useful to gain an intuition for the dynamics observed during the inter-ictal period, as a natural comparator to the ictal state. We therefore apply the dynamical stability-based characterization to four different inter-ictal periods from subject 3. We observe that the angles of the eigenvalues display quite small fluctuations around the mean (Fig.~\ref{fig:SI_Figure4}), in contrast to the large fluctuations observed over the ictal period (see Fig.~\ref{fig:Figure1} and  Fig.~\ref{fig:Figure2}). We do not observe long periods of approximately time-invariant eigenvector loading patterns with strong contribution from only a few regions. Instead, we observe high temporal variance in eigenvector loading patterns, and relatively homogeneous contributions from all electrodes. 

\subsection*{Comparison to clinical annotations}

\begin{figure}
\centering
\includegraphics[width=\linewidth]{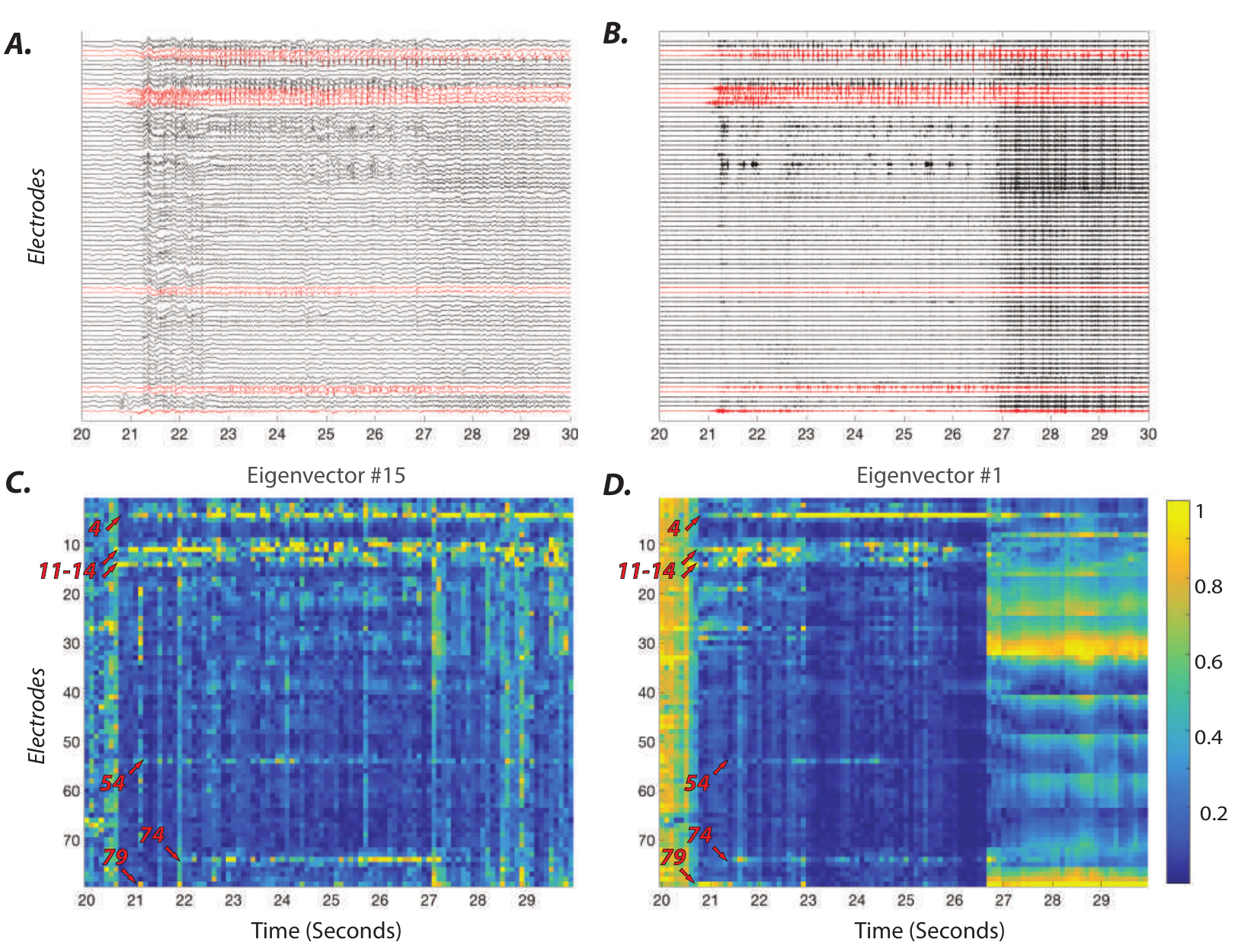}
\caption{\textbf{Evolution of sample eigenvectors of a subject over an ictal period.}  \emph{\textbf{(A)}} ECoG signals from 79 electrodes in subject 3 over a sample ictal period, from 20 s before seizure onset to 20 s after seizure termination, and applying a high-pass filter of $0.1$ Hz. \emph{\textbf{(B)}} The same time-series after applying a high-pass filter at $60$ Hz, highlighting the emergence of high frequency activity at seizure onset. The clinical assessment identifies electrode $79$ as the EEC, marked by a sharp wave followed by low voltage fast activity lasting approximately $3$ s. The EEC is immediately followed by high frequency activity at electrodes $11$-$14$ and later followed by electrodes $3$-$4$, and $53$-$54$ (lasting approximately $7$ s). These time-series are highlighted in red.  \emph{\textbf{(C-D)}} The time evolution of one low and one high frequency eigenvector. Note that the same electrodes identified clinically as the seizure onset zone are also identifiable from the evolution of the eigenvectors (as indicated by red arrows). }
\label{fig:SI_figure_4}
\end{figure}

Seizure onset is marked by the emergence of slow-changing and approximately time-invariant dynamics, commonly with a stereotypical spatial focus. Here, we provide examples that show that these observations accurately echo the clinical annotations, where seizure onset is associated with the emergence of high frequency activity lasting for a few seconds. We begin by studying the evolution of sample eigenmodes of a single patient over the ictal period presented in Fig.~\ref{fig:Figure1}. Here, the clinical assessment of the time-series identifies the EEC at electrode 79, marked by a sharp wave followed by low voltage fast activity lasting $\approx$ 3 s (Fig.~\ref{fig:SI_figure_4}A-B). The clinically-marked EEC is immediately followed by high frequency activity at electrodes 11-14 and later followed by electrodes 3-4, and 53-54 (lasting $\approx$ 7 s).  After applying the dynamical stability-based characterization, we observe that the eigenvectors associated with high frequency activity (the patient's average peri-ictal frequency was 41.91 $\pm$ 15.05 Hz) and middling frequency activity (approximately 29.13 $\pm$ 11.96 Hz) similarly highlight these electrodes (Fig.~\ref{fig:SI_figure_4}C-D). Together, these results show that the assessment of the evolution of the eigenvectors can provide real-time spatiotemporal information about the sources of the slow-changing dynamics associated with seizure onset and termination, which also matches the off-line clinical annotations.

\subsection*{Statistical testing demonstrating spatially constrained and slow changing dynamics}

Here, we provide quantitative analyses supporting the relatively time-invariant nature of seizure onset dynamics in comparison to inter-ictal dynamics. A time-invariant linear system, when estimated using a moving autoregressive model, will ideally yield the same $A$ matrix and, consequently, the very same eigenvectors at every time-point. Nevertheless, in practice, physiological and measurement noise can lead to fluctuations of the estimated system. 

\begin{figure}
\centering
\includegraphics[width=0.85\linewidth]{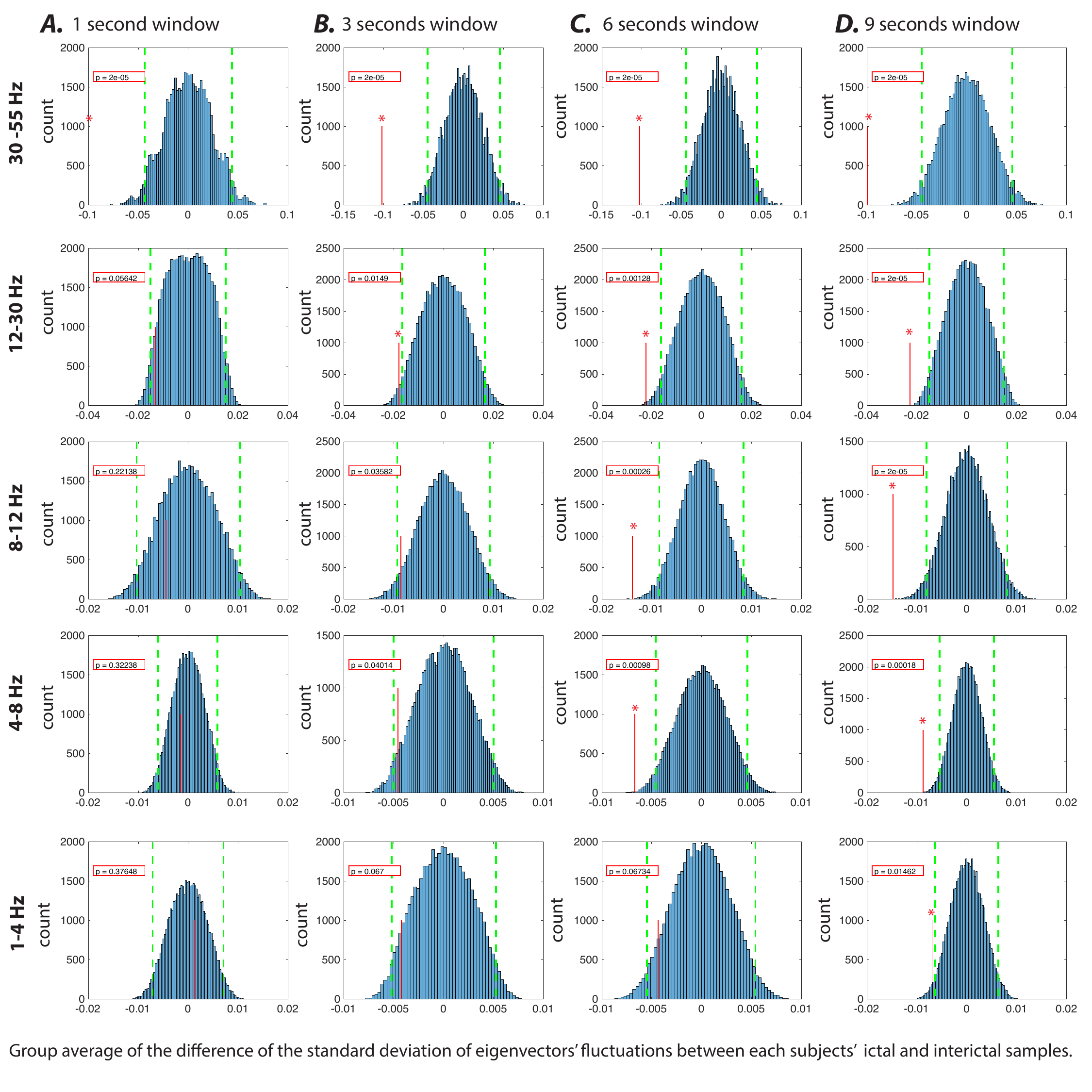}
\caption{\textbf{Approximate spatio-temporal invariance of eigenvector loading patterns at seizure onset.}   
Here we plot the null distributions for the group average of the summary statistic of interest: taking the sum of the element-wise squared differences between temporally contiguous eigenvectors, we then calculated the average standard deviation in this value across all temporally contiguous pairs of eigenvectors across an ictal event.  \emph{\textbf{(A-D)}} Columns indicate different time window sizes (1, 3, 6 and 9 s); rows indicate groupings of eigenmodes into several frequency bands based on their average frequency over samples. The null distribution (shown in blue) was generated by shuffling the ictal and inter-ictal subject-level means and re-calculating the group-level average of subject-level difference between the subjects' ictal and inter-ictal sample means. The dashed green lines show the $95 \%$ confidence interval ($N=50\ 000$ permutations) and the red vertical bar shows the empirical group average of the difference between the subjects' ictal and inter-ictal sample means. Plots with significant results are marked by a red star on the empirical group average. } 
\label{fig:New_SI_Fig_C}
\end{figure}

We first sought to assess the extent of the time-invariance of the eigenvectors over the ictal period at the group-level using non-parametric statistical testing. Specifically, we first defined summary statistic of interest as the sum of the element-wise squared differences between temporally contiguous eigenvectors. Then, we calculated the average standard deviation in this value across all temporally contiguous pairs of eigenvectors across an ictal event. Intuitively, this summary statistic quantifies the temporal variance of the given eigenvector across time: high values of this statistic indicate large temporal fluctuations, and small values of this statistic indicate small temporal fluctuations. To examine this temporal variance at the group level across frequency bands, we grouped the eigenmodes for each patient's dataset based on the average estimated frequency of the eigenmodes into several frequency bands: delta (0-4 Hz), theta (4-8 Hz), alpha (8-12 Hz), beta (12-30 Hz), and gamma (30-55 Hz). We systematically removed the so-called \emph{relaxator} eigenmodes, which are eigenmodes that are non-oscillatory for more than 50 \% of the sample duration. Eigenmodes with average frequencies greater than 55 Hz were removed from the analysis because only a handful of patient recordings displayed this high frequency dynamics. Next, we calculated subject-level summary statistics, averaged over all eigenvectors in each band, for (i) the ictal samples, averaged of all samples for each patient, and (ii) the inter-ictal samples, averaged of all samples for each patient. Then, we compared these summary statistics between ictal and inter-ictal samples at the group-level via bootstrap analysis. To create non-parametric null distributions of the group-level average of the difference between the ictal and inter-ictal summary statistics computed at the subject-level, we permuted the labels ``ictal'' and ``inter-ictal'' uniformly at random 50000 times. Finally, we tested the hypothesis that the eigenvectors in the ictal samples at each frequency band on average displayed significantly less temporal fluctuations than inter-ictal samples across subjects. We observed significantly less eigenvector fluctuations across several frequency bands over the ictal samples in comparison to the inter-ictal samples ($p<0.05$ by permutation testing with $N=50\ 000$; Fig.~\ref{fig:New_SI_Fig_C}). We observe that the difference between the ictal and inter-ictal samples was clear even for the 1 s window following seizure onset in high frequency $\gamma$. However, this effect is best captured with larger window sizes for eigenmodes with slower dynamics. 

\begin{figure}
\centering
\includegraphics[width=0.85\linewidth]{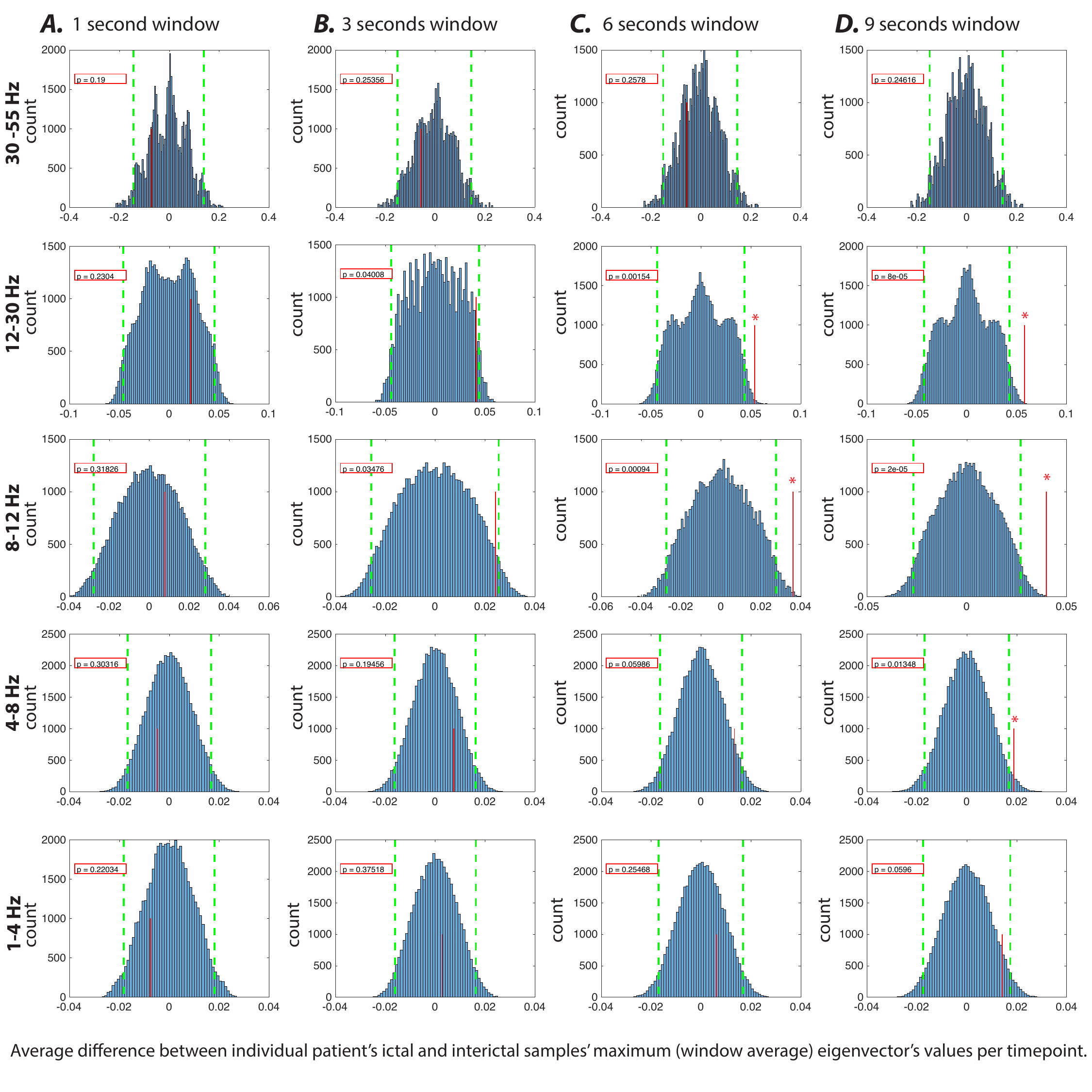}
\caption{\textbf{Increased eigenvector loading following seizure onset compared to the interical period.}  Here we plot the null distributions for the group average obtained as a result of computing the difference between the average maximum entry of the eigenvectors following the seizure onset and that of the inter-ictal samples at the subject-level. \emph{\textbf{(A-D)}} Columns indicate different time window sizes (1, 3, 6 and 9 s); rows indicate groupings of eigenmodes into several frequency bands based on their average frequency over samples. The null distribution was generated by shuffling the ictal and inter-ictal subject-level means and re-calculating the group-level average of subject-level differences between the sample means. The dashed green lines show the $95 \%$ confidence interval ($N=50\ 000$ permutations) and the red vertical bar shows the empirical group average of the difference between the subjects' ictal and inter-ictal sample means. Plots with significant results are marked by a red star on the empirical group average.}
\label{fig:New_SI_Fig_H}
\end{figure}

\begin{figure}
\centering
\includegraphics[width=0.85\linewidth]{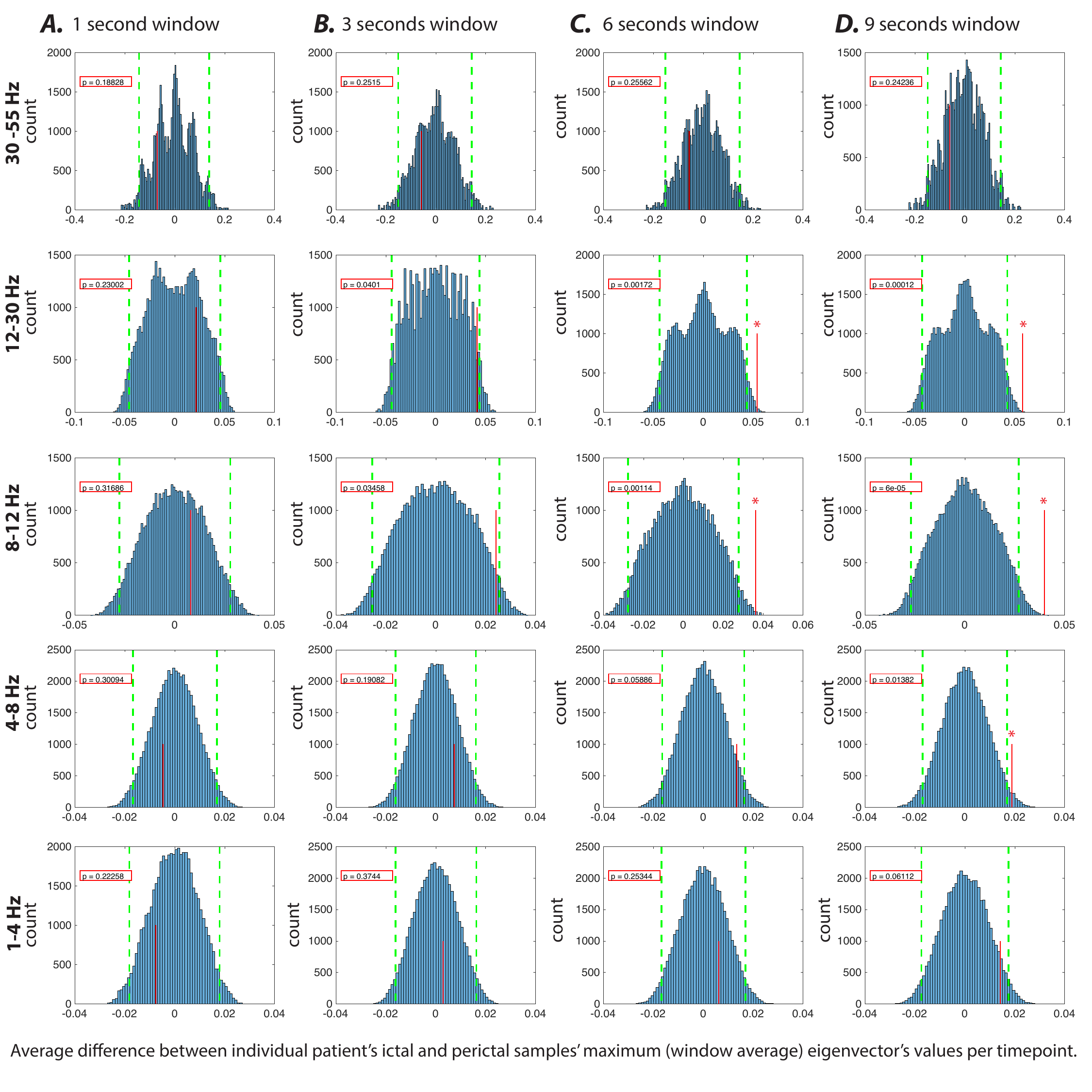}
\caption{\textbf{Increased eigenvector loading following seizure onset compared to the preictal period.}  Here we plot the null distributions for the group average obtained as a result of computing the difference between the average maximum entry of the eigenvectors following the seizure onset and that of the preictal samples at the subject-level. \emph{\textbf{(A-D)}} Columns indicate different time window sizes (1, 3, 6 and 9 s); rows indicate groupings of eigenmodes into several frequency bands based on their average frequency over samples. The null distribution was generated by shuffling the ictal and preictal subject-level means and re-calculating the group-level average of subject-level differences between the sample means. The dashed green lines show the $95 \%$ confidence interval ($N=50\ 000$ permutations) and the red vertical bar shows the empirical group average of the difference between the subjects' ictal and preictal sample means. Plots with significant results are marked by a red star on the empirical group average. Note the similarity between these results and those presented in Fig.~\ref{fig:New_SI_Fig_H}. }
\label{fig:New_SI_Fig_I}
\end{figure}

Following the examination of the time-dependence \emph{versus} time-invariance of eigenvectors during the ictal period, and the frequency dependence of our findings as a function of ictal \emph{versus} inter-ictal samples, we turned to an assessment of the regional localization of the effects. To measure regional localization, we studied the absolute magnitude of the maximally contributing elements of the eigenvectors. Then, we used a bootstrap analysis to test the hypothesis that the emergence of the slow-changing focal ictal dynamics was associated with an increase in the absolute magnitude of the maximally contributing elements of eigenvectors, particularly within a short time window (1, 3, 6, and 9 s) following the seizure onset. We observed systematic increases in the average maximum absolute value of the eigenvector elements following seizure onset across several frequency bands (Fig.~\ref{fig:New_SI_Fig_H}). Interestingly, the high frequency $\gamma$ did not show any increase while $\alpha$ and $\beta$ bands did when considering periods longer than 6 s following the seizure onset. Finally, to determine the specificity of the effects to the ictal period, we compared the subject-level summary statics in the ictal segments to those estimated from the inter-ictal and pre-ictal segments. The comparison between the ictal and preictal samples revealed analogous results as seen in (Fig.~\ref{fig:New_SI_Fig_I}). Together, our complementary analyses reveal the relatively slow-changing nature of seizure onset dynamics across subjects.

\end{document}